\documentclass[aps,prd,onecolumn,groupedaddress,showpacs,nofootinbib,amssymb]{revtex4}
\usepackage[dvips]{graphicx}
\usepackage{amssymb}
\usepackage{diagbox, float}
\usepackage{amsmath}
\usepackage{graphicx,color,epstopdf}
\usepackage{amsfonts}
\usepackage{bm}
\usepackage{cancel}
\usepackage{comment}
\usepackage{longtable}
\usepackage{multirow}

\newcommand{\be}{\begin{equation}}
\newcommand{\ee}{\end{equation}}
\newcommand{\ba}{\begin{eqnarray}}
\newcommand{\ea}{\end{eqnarray}}

\begin{document}
\title{General constraints on Tsallis holographic dark energy from observational data}
\author{Artyom V. Astashenok}
\email{aastashenok@kantiana.ru}
\author{Alexander S. Tepliakov}
\affiliation{I. Kant Baltic Federal University\\
236041, Kaliningrad, Russia\\}%

\begin{abstract}

We investigated Tsallis holographic dark energy (THDE) model in light of modern observations of supernovae, Hubble parameter measurements, data for baryon acoustic oscillations and fluctuations of matter density. The dark energy density for THDE model is written as $\rho_d = 3C^2 / L^{4-2\gamma}$ where $C$ and $\gamma$ are some constants. Scale $L$ is infrared cut-off lenght for which we use the event horizon. For analysis of type Ia supernovae (SNeIa) data Pantheon+ samples are involved. Dark Energy Spectroscopic Instrument (DESI) 2024 measurements serves as source of data about ratios between sound horizon $r_d$ and Hubble ($d_H$) or volume averaged ($d_V$) distances. The updated dataset of Hubble parameter for various redshift is also used in our analysis. Finally we considered the dependence of matter density fluctuations in past from redshift. The standard stratefy of $\chi^2$ minimizing allows to estimate the optimal values of parameters ($\Omega_{de}$ and $H_0$) for some fixed values of $C$ and $\gamma$. One note that best-fit values for parameters $H_0$ from Hubble parameter and SNeIa data are more close than in standard $\Lambda$CDM model for some $C$ and $\gamma$ although problem of Hubble tension remains unsolved. The combined data analysis also gives slightly better results in comparison with standard cosmology. We included in our consideration the possible interaction between matter and holographic component and estimated the acceptable interval of model parameters in this case.

\end{abstract}


\maketitle
\section{Introduction}

Our Universe expands with an acceleration according to observational data concerning type Ia supernovae \cite{1}, \cite{2}. If the universe is filled by only matter and radiation, cosmological acceleration is impossible. Only a substance with negative pressure (``dark energy'') can affect acceleration. As follows from observations this dark energy is distributed in space without inhomogenities and interacts with ordinary matter only by gravitational way.

There have been proposed many models of dark energy. First and most simple theory assumes that dark energy is the vacuum energy (in other words, cosmological contant) $\Lambda$. For cosmological constant pressure is simply $p=-\Lambda$ where $\Lambda$ is density of energy. The corresponding model of Universe is filled with vacuum energy and cold dark matter ($\Lambda$CDM model). This model is described by high accuracy observational data \cite{Amanullah}, \cite{Blake}, \cite{LCDM-1}, \cite{LCDM-2}, \cite{LCDM-3}, \cite{LCDM-4}, \cite{LCDM-5}, \cite{LCDM-6}, \cite{Bamba:2012cp}, \cite{LCDM-7} and therefore can be considered as concordant cosmological model. But one needs to mention the so-called Hubble tension when discussing this model. The best-fit value for current value of Hubble parameter $H_0$ from SNeIa observations significantly differs from $H_0$ determined from Planck mission data. 

Theorethical background of vacuum energy poses additional problems. Firstly, the cosmological constant is very small, although the estimatet vacuum energy value from quantum field theory provides a value that is more hundred orders of magnitude larger than the observed $\Lambda$. Although some assumptions inspired by string theory can reduce this difference (see for example \cite{Linde}), this puzzle remains unresolved. Also, the so called coincidence problem exists: the vacuum energy density is comparable with current matter density in order of magnitude. 

Usually, dark energy is considered as some cosmic fluid. But, there are models of dark energy based on the holographic principle \cite{Wang}, \cite{3}, \cite{4}, \cite{5} from black hole thermodynamics. The holographic principle states that all physical quantities within the universe, including the density of dark energy, can be described by specifying some quantities at the boundary of the universe. Thus, only two physical quantities remain in terms of which the density of dark energy can be expressed: the Planck mass $M_{p}$ and some scale $L$. For $L$, one can take the particle horizon, the event horizon, or the inverse Hubble parameter. The using of the event horizon condition paradoxically assumes that the future evolution of the universe is presupposed, which seems  contradictory from a logical viewpoint. But it is well known that even description of the horizons and interiors of black holes requires boundary conditions in the future \cite{Novik}. The assumption of consistency is similar to ''teleological'' boundary conditions for black holes. For classical holographic dark energy, the energy density in Planck units is $\sim L^{ -2}$.

Various models of holographic dark energy (HDE) satisfy observational constraints \cite{Miao}, \cite{Qing-Guo}, \cite{Qing-Guo-2}. A generalization of a simple HDE model with $\rho_{de}\sim L^{-2}$ is proposed in \cite{Tsallis} (see also \cite{Tsallis-2}). Tsallis holographic dark energy is studied in various papers  (e.g. \cite{Tavayef}, \cite{Jahromi}). Applications of generalized holographic dark energy to modified gravity were considered in \cite{Nojiri-3}. 

Authors of \cite{Nojiri:2021jxf} showed that well-known HDEs including Tsallis, Renyi and Barrow energies are derived from the general form of holographic dark energy proposed in \cite{Nojiri:2019skr}, \cite{Nojiri:2005pu}). In literature various scales $L$ are studied, including particle horizon, Ricci horizon and Grande-Oliveros cutoffs \cite{Zadeh}. Some works (see \cite{Ghaffari}, \cite{Jawad}) are devoted to THDE models in Brans-Dicke gravity and modified Brans-Dicke gravity. Also, consequences of holographic energy on the Randall-Sandrum brane are considered \cite{AA}. A possible interaction between matter and holographic dark energy (\cite{AA2}, \cite{Qihong})  can lead to the disintegration of dark energy and, at a certain intensity, to a quasi-de Sitter expansion of the universe instead of a future singularity.

The current paper is organized the following way. The following section is devoted to the cosmological model with dark energy in form of THDE with scale parameter $L$ defined as length of event horizon. Then we consider various sets of observational data for constraining parameters of the cosmological model. These data include 1701 samples of Ia supernovae from Pantheon+ dataset, updated data of Hubble parameter measurements at various redshifts, DESI-BAO data and data for evolution of matter density fluctuations. In the Section IV we compare THDE model with standard $\Lambda$CDM model. Our analysis assumes fixed parameters of THDE. We found the best-fit values of $\Omega_{de}$ (current fraction of dark energy) and current Hubble parameter $H_0$. Observational datasets are analysed both separately and jointly. From $\chi^2$ statistics it follows that there are such parameters of THDE for which cosmological model describes observational data no worse, and sometimes even a little better, than $\Lambda$CDM model. 

\section{Basic equations}
We consider a spatially flat Universe with the Friedmann-Lemaitre-Robertson-Walker (FLRW) metric:
\begin{equation} 
\label{eq:1}
ds^2=dt^2-a^2(t)(dx^2 + dy^2 + dz^2)
\end{equation}
where $t$ is the cosmological time, $a(t)$ is the scale factor. Assume that the universe is filled with dark energy, matter and radiation with densities $\rho_{de}$, $\rho_m$ and $\rho_{r}$ respectively. The cosmological equations for FLRW metric can be written in the following form:
\begin{equation}
\label{eq:2}
H^{2} = \frac{1}{3}(\rho_{m}+\rho_{de}+\rho_r),
\end{equation}
\begin{equation}
\label{eq:3}
\dot{H} = -\frac{1}{2}(\rho_{m}+\rho_{de}+4\rho_r/3+p_{de}).
\end{equation}
Here $\rho_m$, $\rho_r$ and $\rho_{de}$ are densities of matter, radiation and dark energy correspondingly. The Hubble parameter is defined as $H=\dot{a}/a$. In the Tsallis model, the density of holographic dark energy is
\begin{equation}
\label{eq:4}
\rho_{de}=\frac{3C^{2}}{L^{4-2\gamma }}.
\end{equation}
Here $C$ and $\gamma$ are some constants. For $\gamma = 1$ we have a simple model of holographic dark energy,  $\gamma=2$ corresponds to the vacuum energy. As the scale $L$ consider the future event horizon:
$$L = a \int_{t}^{\infty}\frac{d{t}'}{a}.$$
If dark energy and matter interact with each other, then the continuity equations for the corresponding components take the form:
\begin{equation}
\label{eq:5}
\dot{\rho}_{m}+3H\rho_{m}=-Q,
\end{equation}
\begin{equation}
\label{eq:6}
\dot{\rho}_{de}+3H(\rho_{de}+p_{de})=Q.
\end{equation}
Here, a function $Q$ describing the interaction is introduced. In general case, the value of $Q$ depends on time and densities $\rho_{de}$, $\rho_m$. The total density $\rho=\rho_{de}+\rho_m + \rho_r$ satisfies the usual continuity equation. We can obtain an expression for the pressure of dark energy $p_{de}$.
\begin{equation}
\label{eq:7}
p_{de}=-\frac{\dot{\rho}_{de}}{3H}-\rho_{de}+\frac{Q}{3H}.
\end{equation}
Then the parameter of the equation of state for dark energy is
\begin{equation}
\label{eq:8}
w_{de}=\frac{p_{de}}{\rho_{de}}=-1-\frac{\dot{\rho}_{de}}{3H\rho_{de}}-\frac{Q}{3H\rho_{de}}
\end{equation}

The holographic dark energy model with an event horizon as infrared cut-off can be consistent with observational data for various values of $\gamma$ and $C$, which makes this model quite attractive for explaining of the cosmological acceleration. Now we consider various sets of observational data for determining the best-fit model parameters such as $H_0$ and $\Omega_{de}$. Our strategy consists of defining $\gamma$ and $C$ and finding the best-fit values of $H_0$ and $\Omega_{de}$ by chi-square minimization. For simplicity, we don't consider the contribution of radiation into total energy budget because this contribution is negligible for considered redshifts. 

\section{Observational data}

The observational data involved in the conducted analysis includes the following probes.

\subsection{SNeIa test}  

The relation between SNeIa's distance modulus and redshift are often used to constrain the parameters of dark energy model. The theoretical value of distance modulus for some supernova can be calculated as
\be
  \mbox{\bf $\mu$} = 5 \log_{10}\left[\frac{d_L(z_{hel},z_{cmb})}{Mpc}\right] + 25,
\ee
where $z_{cmb}$ and $z_{hel}$ are the cosmic microwave background restframe and heliocentric redshifts of supernova. The luminosity distance ${d}_L$ is given by
\be
  {d}_L(z_{hel},z_{cmb}) = (1+z_{hel}) r(z_{cmb}),
\ee
where corresponding comoving distance $r(z)$ is
\be
\label{eq:rz}
r(z) = H_0^{-1}\, |\Omega_k|^{-1/2} {\rm sinn}\left[|\Omega_k|^{1/2}\, \int_0^z\frac{dz'}{E(z')}\right].
\ee
Here $E(z) \equiv H(z)/H_0$ is the dimensionless Hubble parameter,
${\rm sinn}(x)=\sin(x)$, $x$, $\sinh(x)$ for negative curvature ($\Omega_k<0$), flat space ($\Omega_k=0$), and positive curvature ($\Omega_k>0$), respectively. We will restrict ourselves to the case of flat spacetime. 

Pantheon+ is one of the recent and detailed catalogues of SNe I. It contains 1701 sources \cite{Torny}, \cite{Skolnik}, \cite{Brout}. The general best-fit model parameters correspond to maximum of the total log-likelihood function
$$
\ln \mathcal{L}_{SNe} = -\frac{1}{2}\left(\Delta {\bf \mu}^T \cdot \mbox{\bf C}^{-1}_{SNe} \cdot \Delta{\bf \mu} + \ln (2\pi |{\bf C}_{SNe}|\right)),
$$
where the i-th component of the vector $\Delta \mu$ is
$$
\Delta\mu_{i}\equiv \mu_{i}-\mu(z_{ihel}, z_{icmb}).
$$
The total covariance matrix $\mbox{\bf C}_{SNe}$ can be expressed as
\be
\mbox{\bf C}_{SNe}=\mbox{\bf D}_{\rm stat}+\mbox{\bf C}_{\rm stat}
+\mbox{\bf C}_{\rm sys}.
\ee
Here $\mbox{\bf D}_{\rm stat}$ is the diagonal part of the statistical
uncertainty, which is given by \cite{Skolnik},
\ba
\mbox{\bf D}_{\rm stat,ii}&=&\left[\frac{5}{z_i \ln 10}\right]^2 \sigma^2_{z,i}+
  \sigma^2_{\rm int} +\sigma^2_{\rm lensing} + \sigma^2_{m_B,i} \nonumber\\
&&   +\alpha^2 \sigma^2_{X_1,i}+\beta^2 \sigma^2_{{\cal C},i}
+ 2 \alpha C_{m_B X_1,i} - 2 \beta C_{m_B {\cal C},i} -2\alpha\beta C_{X_1 {\cal C},i}.
\ea
The first three terms arise due to the uncertainty in redshift as a result of peculiar velocities,
the intrinsic variation in supernova magnitude, and the variation of magnitudes caused by gravitational lensing. Errors $\sigma^2_{m_B,i}$, $\sigma^2_{X_1,i}$, and $\sigma^2_{{\cal C},i}$
denote the uncertainties of $m_B$, $X_1$ and ${\cal C}$ for the $i$-th SN.
In addition, $C_{m_B X_1,i}$, $C_{m_B {\cal C},i}$ and $C_{X_1 {\cal C},i}$
are the covariances between $m_B$, $X_1$ and ${\cal C}$ for the $i$-th supernova.
Finally $\mbox{\bf C}_{\rm stat}$ and $\mbox{\bf C}_{\rm sys}$
are the statistical and the systematic covariance matrices, given by
\be
\mbox{\bf C}_{\rm stat}+\mbox{\bf C}_{\rm sys}=V_0+\alpha^2 V_a + \beta^2 V_b +
2 \alpha V_{0a} -2 \beta V_{0b} - 2 \alpha\beta V_{ab},
\ee
where $V_0$, $V_{a}$, $V_{b}$, $V_{0a}$, $V_{0b}$ and $V_{ab}$ are six $1701\times1701$ matrices. We use statistical and systematical part of covariance matrix which given on cite \cite{Pantheon_SH0}. The errors to value of distance modulus are also given in table from this source.  

Let us introduce vector 
$$
\mu^{(0)} = \mu - \mu_{0}, \quad \mu_0 =  5\log_{10}\left(\frac{d_{H_{0}}}{Mpc}\right), 
$$
$$
d_{H_{0}} = \frac{c}{H_0}
$$
The value of $\chi^2$ for Pantheon+ dataset is
\be  \label{eq:chi2_SN}
  \chi^2_{SNe} = A_{SNe} - 2\mu_0 B_{SNe} + \mu_{0}^{2} S_{SNe},  
\ee
where the values of $A_{SNe}$, $B_{SNe}$ and $S_{SNe}$ are defined as
$$
A_{SNe} = (\mu_{obs}-\mu^{(0)})^{T}\cdot C^{-1}_{SNe} \cdot (\mu_{obs}-\mu^{(0)}),
$$
$$
B_{SNe} = (\mu_{obs} - \mu^{(0)})^{T}\cdot C^{-1}_{SNe} I, \quad S_{SNe} = I^TC^{-1}_{SNe}I, 
$$
where $I^T = (1,1,...1)$. The $\chi^2_{SNe}$ has a minimum for any fixed parameters
$$
\bar{\chi}^{2}_{SNe}= A_{SNe} - B_{SNe}^2 / S_{SNe}.
$$
if 
$$
\mu_{0}^{(opt)} = B_{SNe}/S_{SNe}.
$$
Therefore  for given $\Omega_{de}$ we can find the best-fit value of $H_0$ for Pantheon+ data. This approach is used when likelihood for $H_0$ is averaged on $\Omega_{de}$.

\subsection{BAO test}

Baryon acoustic oscillations (BAO) \cite{Blake} give the value of acoustic parameter $A$ as function of redshift. The corresponding theoretical value of this parameter is
\begin{equation}\label{Ath}
A_{th}(z)=\frac{d_{V}(z)H_{0}\sqrt{\Omega_{m}}}{z},
\end{equation}
where $d_{V}(z)$ is a distance parameter defined as
\begin{equation} \label{DV}
d_{V}(z)=\left\{d_{L}^{2}(z)\frac{cz}{(1+z)^{2}H(z)}\right\}^{1/3}.
\end{equation}
Here $c$ is the sound speed in epoch of recombination. There are three types of BAO measurements. Firstly, relation between Hubble distance and the sound horizon at the drag epoch, $r_d$,
$$
\frac{d_H(z)}{r_d}=\frac{c\,r_d^{-1}}{H(z)}.
$$
Then  
$$
\frac{d_V(z)}{r_d}= \left[\frac{c\,z\,r_d^{-3} d_L^2(z)}{H(z)(1+z)^2}\right]^{\frac{1}{3}}
$$
and finally relation $r(z)/r_d$. For first two types of measurements we take data from galaxy surveys spanning a redshift range $z\in[0.1,4.2]$ and divided into $N_{B}=7$ distinct redshift bins \cite{DESI:2024mwx}. For five values of $z_{eff}$ we use also measurements of relation $r(z)/r_d$. As noted in \cite{DESI:2024mwx}, these data are effectively independent from each other and, thus, no covariance matrix is considered. 

BAO measurements data are insensitive to value $H_0 r_d$. The $r_d$ is a function of the matter and baryon densities and the effective number of extra-relativistic degrees of freedom. For illustration we use fixed value of $r_d$ in our calculation, namely $147.09$ Mpc. This value falls within both the Planck satellite and DESI-BAO expectations.

\begin{table}[ht]
\centering
\setlength{\tabcolsep}{1.em}
\renewcommand{\arraystretch}{1.1}
\begin{tabular}{|l|c|c|c|c|}
\hline
Tracer  & $z_{\rm eff}$ & $d_H/r_d$         & $d_V/r_d$     & $r/r_d$ \\
\hline
BGS     & $0.30$        & $-$               & $7.93\pm0.15$ & \\
LRG     & $0.51$        & $20.98\pm0.61$    & $-$           & $13.62\pm 0.25$\\
LRG     & $0.71$        & $20.08\pm0.60$    & $-$           & $16.85\pm 0.32$\\
LRG+ELG & $0.93$        & $17.88\pm0.35$    & $-$           & $21.71\pm 0.28$ \\
ELG     & $1.32$        & $13.82\pm0.42$    & $-$           & $27.79\pm 0.69$ \\
QSO     & $1.49$        & $-$               & $26.07\pm0.67$ & \\
Lya QSO & $2.33$        & $8.52\pm0.17$     & $-$            & $39.71\pm 0.94$ \\
\hline
\end{tabular}
\caption{DESI-BAO data with tracers, effective redshifts $z_{\rm eff}$, and ratios $d_H/r_d$, $d_V/r_d$ and $r/r_d$. Data are taken from \cite{DESI:2024mwx}.}
\label{tab:BAO}
\end{table}
For all three types of these measurements $X_i$ we calculate corresponding log-likelihood function
\begin{equation}
\label{loglikeBAOu}
    \ln \mathcal{L}_{\rm X} = -\frac{1}{2} \sum_{i=1}^{N_{\rm X}}\left\{\left[\dfrac{X_i-X(z_i)}{\sigma_{X_i}}\right]^2 + \ln(2\pi\sigma^2_{X_j})\right\},
\end{equation}
and the total BAO log-likelihood is
\begin{equation}
\label{loglikeBAO}
    \ln \mathcal{L}_{\rm BAO} = \sum_X\ln \mathcal{L}_{\rm X}\,.
\end{equation}
We define $\chi^{2}_{BAO}$ as 
$$
\chi^{2}_{BAO} = \sum_{i=1}^{N_{\rm X}}\left[\dfrac{X_i-X(z_i)}{\sigma_{X_i}}\right]^2
$$
or
\begin{equation}
   \chi^{2}_{BAO} = A_{BAO} - 2B_{BAO}\nu + \nu^2 S_{BAO}, \quad \nu = \frac{1}{r_d H_0},
\end{equation}
where
$$
A_{BAO} = \sum_{i=1}^{N_{\rm X}}\left(\dfrac{X_i}{\sigma_{X_i}}\right)^2, \quad B_{BAO} = \sum_{i=1}^{N_{\rm X}}\dfrac{X_i X^{(0)}(z_i)}{\sigma_{X_i}^{2}}, \quad S_{BAO}=\sum_{i=1}^{N_{\rm X}}\left(\dfrac{X^{(0)}(z_i)}{\sigma_{X_i}}\right)^2.
$$
$$
X^{(0)}(z_i) = r_d H_{0} X(z_i).
$$
Similarly to the case of Pantheon+ dataset described above, we can find the best-fit value of $H_{0}$ for given $\Omega_{de}$:
$$
\nu^{(opt)} = B_{BAO}/S_{BAO}.
$$

\subsection{Hubble parameter test} 

The most recent dataset of Hubble parameter as function of redshift consists of $36$ measurements (see Table~\ref{tab:OHD}). Hubble parameter can be determined due to differences between age and redshift of couples of passively evolving galaxies. Assuming that considered galaxies are formed at the same time, we conclude that $H(z)=-(1+z)^{-1}\Delta z/\Delta t$ \cite{Jimenez:2001gg}.

However the errors of these measurements can be considerable, because they strongly depend on various assumptions about stellar population synthesis. Uncertainty of initial mass functions and the stellar metallicity may contribute at most to $20$--$30\%$ errors  \cite{Montiel:2020rnd,Moresco:2022phi,Rom:2023kqm}. 

Assuming gaussian distributed errors $\sigma_{H_k}$, we can found the best-fit parameters by maximizing the log-likelihood function
\begin{equation}
\label{loglikeOHD}
    \ln \mathcal{L}_{\rm H} = -\frac{1}{2} \sum_{i=1}^{36}\left\{\left[\dfrac{H_i-H(z_i)}{\sigma_{H_i}}\right]^2 + \ln(2\pi\sigma^2_{H_i})\right\}\,.
\end{equation}
Again, as in the case of SNe and BAO data, we can perform minimization over possible values of $H_0$ for given $\Omega_{de}$. For $\chi^2_{H}$ we have relation
$$
\chi^{2}_{H}=A_{H}-2B_{H}H_{0}+H_{0}^{2}S_{H},
$$
$$
A_{H}=\sum_{i}\frac{H_{i}^{2}}{\sigma^{2}_{H_i}},\quad B_{H}=\sum_{i}\frac{E(z_{i})H_{i}}{\sigma^{2}_{H_i}},\quad
$$
$$
S_{H}=\sum_{i}\frac{E^{2}(z_i)}{\sigma^{2}_{H_i}}.
$$
The parameter $\chi^{2}_{H}$ has a minimum for $H_{0}^{(opt)}=\sqrt{B_{H}/S_{H}}$ and this minimum is
$$
{\bar{\chi}}_{H}^{2}=A_{H}-B_{H}^{2}/S_{H}.
$$

\begin{table}[ht]
\centering
\setlength{\tabcolsep}{1.5em}
\renewcommand{\arraystretch}{1.1}
\begin{tabular}{|c|c|c|c|}
   \hline
   Index & $z$     &$H(z),$ &  Ref. \\
          &  &km/s/Mpc&\\
    \hline
1 &    0.0708  & $69.0\pm 19.6\pm12.4^\star$ & \cite{Zhang:2012mp} \\
2 &   0.09    & $69.0 \pm12.0\pm11.4^\star$  & \cite{Jimenez:2001gg} \\
3 &    0.12    & $68.6\pm26.2\pm11.4^\star$  & \cite{Zhang:2012mp} \\
4 &    0.17    & $83.0\pm8.0\pm13.1^\star$   & \cite{Simon:2004tf} \\
5 &    0.1791   & $75.0  \pm 3.8\pm0.5^\dagger$   & \cite{Moresco:2012jh} \\
6 &    0.1993   & $75.0\pm4.9\pm0.6^\dagger$   & \cite{Moresco:2012jh} \\
7 &    0.20    & $72.9\pm29.6\pm11.5^\star$  & \cite{Zhang:2012mp} \\
8 &    0.240	& $79.69	\pm 2.65$ & \\
9 &   0.27    & $77.0\pm14.0\pm12.1^\star$  & \cite{Simon:2004tf} \\
10 &    0.28    & $88.8\pm36.6\pm13.2^\star$  & \cite{Zhang:2012mp} \\
11 &    0.3519   & $83.0\pm13.0\pm4.8^\dagger$  & \cite{Moresco:2016mzx} \\
12 &    0.3802  & $83.0\pm4.3\pm12.9^\dagger$  & \cite{Moresco:2016mzx} \\
13 &    0.4     & $95.0\pm17.0\pm12.7^\star$  & \cite{Simon:2004tf} \\
14 &    0.4004  & $77.0\pm2.1\pm10.0^\dagger$  & \cite{Moresco:2016mzx} \\
15 &    0.4247  & $87.1\pm2.4\pm11.0^\dagger$  & \cite{Moresco:2016mzx} \\
16 &    0.430	& $86.45\pm 3.68$    & \\
17   & 0.4497  & $92.8 \pm4.5\pm 12.1^\dagger$  & \cite{Moresco:2016mzx} \\
18   & 0.47    & $89.0\pm23.0\pm44.0^\dagger$     & \cite{Ratsimbazafy:2017vga}\\
 19  & 0.4783  & $80.9\pm2.1\pm 8.8^\dagger$   & \cite{Moresco:2016mzx} \\
20   & 0.48    & $97.0\pm62.0\pm12.7^\star$  & \cite{Stern:2009ep} \\
21   & 0.5929   & $104.0\pm11.6\pm4.5^\dagger$  & \cite{Moresco:2012jh} \\
22   & 0.6797    & $92.0\pm6.4\pm4.3^\dagger$   & \cite{Moresco:2012jh} \\
23   & 0.75    & $98.8\pm24.8\pm22.7^\dagger$     & \cite{Borghi:2021rft}\\
24   & 0.7812   & $105.0\pm9.4\pm6.1^\dagger$  & \cite{Moresco:2012jh} \\
25   & 0.80    & $113.1\pm15.1\pm20.2^\star$    & \cite{Jiao:2022aep}\\
26   & 0.8754   & $125.0\pm15.3\pm6.0^\dagger$  & \cite{Moresco:2012jh} \\
27   & 0.88    & $90.0\pm40.0\pm10.1^\star$  & \cite{Stern:2009ep} \\
28   & 0.9     & $117.0\pm23.0\pm13.1^\star$  & \cite{Simon:2004tf} \\
29   & 1.037   & $154.0\pm13.6\pm14.9^\dagger$  & \cite{Moresco:2012jh} \\
30   & 1.26    & $135.0\pm60.0\pm27.0^\dagger$   & \cite{Tomasetti:2023kek} \\
31   & 1.3     & $168.0\pm17.0\pm14.0^\star$  & \cite{Simon:2004tf} \\
32   & 1.363   & $160.0 \pm 33.6^\ddag$  & \cite{Moresco:2015cya} \\
33   & 1.43    & $177.0\pm18.0\pm14.8^\star$  & \cite{Simon:2004tf} \\
34   & 1.53    & $140.0\pm14.0\pm11.7^\star$  & \cite{Simon:2004tf} \\
35   & 1.75    & $202.0\pm40.0\pm16.9^\star$  & \cite{Simon:2004tf} \\
36   & 1.965   & $186.5 \pm 50.4^\ddag$  & \cite{Moresco:2015cya} \\
\hline
\end{tabular}
\caption{Redshifts, Hubble parameter measurements with errors and corresponding references. Systematic errors computed here are labeled with $\star$, with $\dagger$ if given by the literature, and with $\ddag$ when combined with statistical errors.}
\label{tab:OHD}
\end{table}
\subsection{Matter density perturbations} 

We consider the theory of linear perturbations in the frames of holographic dark energy model by assuming scalar fluctuations of the metric in the Newtonian gauge \cite{Mukhanov92}:
\begin{equation}
ds^2=-(1+2\phi)dt^2+a^2(t)(1-2\phi)\delta_{ij}dx^idx^j\ ,
\end{equation}
where $\phi$ is scalar perturbation. We consider a simplest scenario with homogeneous holographic dark energy ($\delta \rho_{de}=0$) and therefore clustering occurs due only to the corresponding matter component.
For density contrast of matter $\delta_m \equiv \delta\rho_m/\rho_m$ and the divergence of the fluid velocity $\theta_m\equiv \vec{\nabla}\cdot \vec{v}_m$ the equations should be satisfied in the Fourier space:
\begin{align}
&\dot{\delta}_m+\dfrac{\theta_m}{a}-3\dot{\phi}=0\ , \label{eq:deltamdot}\\
&\dot{\theta}_m+H\theta_m-\dfrac{k^2\phi}{a}=0\ , \label{eq:thetamdot} \\
\end{align}
where potential $\phi$ is governed by equation
\begin{equation}
    \ddot{\phi}+4H\dot{\phi}+\left(2\dfrac{\ddot{a}}{a}+H^2\right)\phi=0.
\end{equation}
We analyze the case of a matter-dominated universe, where $\phi$ is a constant. For potential $\phi$ at sub-horizon scales $(k^2\gg a^2H^2)$ the Poisson equation in Fouries space turns into
\begin{equation}
\frac{k^2}{a^2}\phi=\dfrac{3}{2}H^2 \Omega_m(a)\delta_m.
\end{equation}
Here $\Omega_m(a)$ is the fraction of matter density in total density as function of scale factor. Therefore we obtain the evolution equations for the dark matter perturbations \cite{Tsujikawa:2007gd,Nesseris:2015fqa,Arjona:2018jhh}:
\begin{equation}
\delta''_m+A_m\delta'_m+S_m\delta_m = 0, \label{eq:matter}
\end{equation}
where 
$$
A_m=\dfrac{3}{2a}(1-w_{de}\Omega_{de}(a)), \quad S_m=-\dfrac{3}{2a^2}\Omega_{m}(a).
$$
For initial conditions we take $\delta_{m}(a_{in}) = a_{in}$, $\delta'_{m}(a_{in}) = 1$  and start integration from $a_{in} = 0.01$. The quantity $f(a)\equiv a \delta'_m(a)$ defines the growth rate of density perturbations. From redshift-space distortion measurements in the interval $0<z<2$ the factor
\begin{equation}
f\sigma_8(a)\equiv f(a)\sigma_8(a)\ 
\end{equation} 
is available (see Table III). The $\sigma_8(a)=\sigma_8\delta_m(a)/\delta(1)$ estimates the linear-density field fluctuations within a $8h^{-1}\text{Mpc}$ radius, with $\sigma_8$ being its current value.
For current cosmology we can calculate evolution of matter perturbation and compare results with observations. One needs, however, to also rescale the results with respect to the assumed fiducial cosmology. In order to do that we introduce the fiducial Alcock-Paczynski correction factor $q(z)$ \cite{Macaulay:2013swa,Kazantzidis:2018rnb} 
\begin{equation}
q(z)=\dfrac{H(z)d_A(z)}{H_{fid}(z)d_{A,fid}(z)}
\label{correction}
\end{equation}
to convert angles and redshift to distances for evaluating the correlation function. The subscript `\emph{fid}' refers to the fiducial $\Lambda$CDM model characterized by the following dependence of Hubble parameter from redshift:
\begin{equation}
H_{fid}(z)=H_0\sqrt{\Omega_{m}(1+z)^3+1-\Omega_{m}}.
\end{equation}
$d_A(z)$ is the the angular diameter distance. In terms of redshift and dimensionless Hubble parameter, Eq. (\ref{eq:matter}) takes the form \cite{EspositoFarese:2000ij}
\begin{equation}
\frac{d^2\delta_m}{dz^2} + \left(\frac{1}{E}\frac{dE}{dz} -
\frac{1}{1+z}\right)\frac{d\delta_{m}}{dz}
-\frac{3}{2} \frac{\Omega_{m}(z)
}{(1+z)^2E^2}~\delta_m=0.
\label{eq:odedeltaz}
\end{equation}
Once the evolution of $\delta$ is known, the observable product $f\sigma_8(a)\equiv f(a)\cdot \sigma(a)$  can be obtained using the definitions. Thus, we have 
\be
f\sigma_8(a,\sigma_8,\Omega_{m})=\frac{\sigma_8}{\delta_m(1)}~a~\delta'_m(a) .
\label{eq:fs8}
\ee
This theoretical prediction may now be used to be compared with the observed $f\sigma_8(z)$ data and to obtain fits for the parameters $\Omega_{m}$, $\sigma_8$. The $f\sigma_8$ data are not depend from $H_0$. For the construction of $\chi^2_{f\sigma_8}$ we use the vector \cite{Kazantzidis:2018rnb}
\be 
V_{f\sigma_8}^i(z_i)\equiv f\sigma_{8,i}^{obs}-\frac{f\sigma_8^{th}(z_i)}{q(z_i)}
\ee 
where $f\sigma_{8,i}^{obs}$ is the the value of the $i$th datapoint, with $i= 1,...,N_{f\sigma_8}$ (where $N_{f\sigma_8}=66$ corresponds to the total number of datapoints of Table \ref{tab:data-rsd}) and $f\sigma_8^{th}(z_i)$ is the theoretical prediction, both at  redshift $z_i$.  

Thus, we obtain corresponding $\chi_{f\sigma_8}^2$ as  
\be
\chi_{f\sigma_8}^2(\Omega_{m},\sigma_8)=V^{T}_{f\sigma_8}C_{f\sigma_8}^{-1}V_{f\sigma_8}
\label{chif}
\ee
where $C_{f\sigma_8}^{-1}$ is  the  inverse matrix  to the covariance matrix which is assumed to be diagonal with the exception of the $3\times3$ WiggleZ subspace (see \cite{Kazantzidis:2018rnb} for more details about this compilation).

\begin{longtable}[t]{|c|c|c|c|c|}
\hline
    Index & Dataset & $z$ & $f\sigma_8(z)$  & fiducial cosmology \\ 
\hline
1 & SDSS-LRG & $0.35$ & $0.440\pm 0.050$   &$(\Omega_{m},\sigma_8$)$=(0.25,0.756)$\cite{Tegmark:2006az} \\

2 & VVDS & $0.77$ & $0.490\pm 0.18$  & $(\Omega_{m},\sigma_8)=(0.25,0.78)$ \\

3 & 2dFGRS & $0.17$ & $0.510\pm 0.060$    & $(\Omega_{m}, \sigma_8)=(0.3, 0.9)$ \\

4 &2MRS &0.02& $0.314 \pm 0.048$  & $(\Omega_{m},\sigma_8)=(0.266,0.65)$ \\

5 & SnIa+IRAS &0.02& $0.398 \pm 0.065$  & $(\Omega_{m},\sigma_8)=(0.3,0.814)$\\

6 & SDSS-LRG-200 & $0.25$ & $0.3512\pm 0.0583$   & $(\Omega_{m},\sigma_8)=(0.276,0.8)$  \\

7 & SDSS-LRG-200 & $0.37$ & $0.4602\pm 0.0378$   & \\

8 & SDSS-LRG-60 & $0.25$ & $0.3665\pm0.0601$   &  \\

9 & SDSS-LRG-60 & $0.37$ & $0.4031\pm0.0586$   &\\

10 & WiggleZ & $0.44$ & $0.413\pm 0.080$   & $(\Omega_{m},\sigma_8)=(0.27,0.8)$ \\

11 & WiggleZ & $0.60$ & $0.390\pm 0.063$  &  \\

12 & WiggleZ & $0.73$ & $0.437\pm 0.072$  &\\

13 & 6dFGS& $0.067$ & $0.423\pm 0.055$  & $(\Omega_{m},\sigma_8)=(0.27,0.76)$ \\

14 & SDSS-BOSS& $0.30$ & $0.407\pm 0.055$   & $(\Omega_{m},\sigma_8)=(0.25,0.804)$ \\

15 & SDSS-BOSS& $0.40$ & $0.419\pm 0.041$   & \\

16 & SDSS-BOSS& $0.50$ & $0.427\pm 0.043$   & \\

17 & SDSS-BOSS& $0.60$ & $0.433\pm 0.067$  & \\

18 & VIPERS& $0.80$ & $0.470\pm 0.080$  & $(\Omega_{m},\sigma_8)=(0.25,0.82)$  \\

19 & SDSS-DR7-LRG & $0.35$ & $0.429\pm 0.089$   & $(\Omega_{m},\sigma_8$)$=(0.25,0.809)$\cite{Komatsu:2010fb}\\

20 & GAMA & $0.18$ & $0.360\pm 0.090$    & $(\Omega_{m},\sigma_8)=(0.27,0.8)$ \\

21& GAMA & $0.38$ & $0.440\pm 0.060$   & \\

22 & BOSS-LOWZ& $0.32$ & $0.384\pm 0.095$    & $(\Omega_{m},\sigma_8)=(0.274,0.8)$ \\

23 & SDSS DR10+DR11 & $0.32$ & $0.48 \pm 0.10$   & $(\Omega_{m},\sigma_8$)$=(0.274,0.8)$\cite{Anderson:2013zyy}\\

24 & SDSS DR10+DR11 & $0.57$ & $0.417 \pm 0.045$   &  \\

25 & SDSS-MGS & $0.15$ & $0.490\pm0.145$   & $(\Omega_{m},\sigma_8)=(0.31,0.83)$ \\

26 & SDSS-veloc & $0.10$ & $0.370\pm 0.130$   & $(\Omega_{m},\sigma_8$)$=(0.3,0.89)$\cite{Tegmark:2003uf} \\

27 & FastSound& $1.40$ & $0.482\pm 0.116$   & $(\Omega_{m},\sigma_8$)$=(0.27,0.82)$\cite{Hinshaw:2012aka} \\

28 & SDSS-CMASS & $0.59$ & $0.488\pm 0.060$  & $(\Omega_{m},\sigma_8)=(0.307,0.8288)$ \\

29 & BOSS DR12 & $0.38$ & $0.497\pm 0.045$   & $(\Omega_{m},\sigma_8)=(0.31,0.8)$ \\

30 & BOSS DR12 & $0.51$ & $0.458\pm 0.038$   & \\

31 & BOSS DR12 & $0.61$ & $0.436\pm 0.034$   & \\

32 & BOSS DR12 & $0.38$ & $0.477 \pm 0.051$  &  \\

33 & BOSS DR12 & $0.51$ & $0.453 \pm 0.050$   & \\

34 & BOSS DR12 & $0.61$ & $0.410 \pm 0.044$  &  \\

35 &VIPERS v7& $0.76$ & $0.440\pm 0.040$   & $(\Omega_{m},\sigma_8)=(0.308,0.8149)$ \\

36 &VIPERS v7 & $1.05$ & $0.280\pm 0.080$  &\\

37 &  BOSS LOWZ & $0.32$ & $0.427\pm 0.056$   & $(\Omega_{m},\sigma_8)=(0.31,0.8475)$\\

38 & BOSS CMASS & $0.57$ & $0.426\pm 0.029$  & \\

39 & VIPERS  & $0.727$ & $0.296 \pm 0.0765$  & $(\Omega_{m},\sigma_8)=(0.31,0.7)$\\

40 & 6dFGS+SnIa & $0.02$ & $0.428\pm 0.0465$   & $(\Omega_{m},\sigma_8)=(0.3,0.8)$ \\

41 &VIPERS PDR2& $0.60$ & $0.550\pm 0.120$   & $(\Omega_{m},\sigma_8)=(0.3,0.823)$ \\

42 & VIPERS PDR2& $0.86$ & $0.400\pm 0.110$   & \\

43 & SDSS DR13  & $0.1$ & $0.48 \pm 0.16$   & $(\Omega_{m},\sigma_8$)$=(0.25,0.89)$\cite{Tegmark:2003uf} \\

44 & 2MTF & 0.001 & $0.505 \pm 0.085$   & $(\Omega_{m},\sigma_8)=(0.3121,0.815)$ \\

45 & VIPERS PDR2 & $0.85$ & $0.45 \pm 0.11$   &  $\Omega_{m}=0.30$ \\

46 & BOSS DR12 & $0.31$ & $0.384 \pm 0.083$ & $(\Omega_{m},\sigma_8)=(0.307,0.8288)$\\

47 & BOSS DR12 & $0.36$ & $0.409 \pm 0.098$   & \\

48 & BOSS DR12 & $0.40$ & $0.461 \pm 0.086$   & \\

49 & BOSS DR12 & $0.44$ & $0.426 \pm 0.062$   & \\

50 & BOSS DR12 & $0.48$ & $0.458 \pm 0.063$   & \\

51 & BOSS DR12 & $0.52$ & $0.483 \pm 0.075$  & \\

52 & BOSS DR12 & $0.56$ & $0.472 \pm 0.063$   & \\

53 & BOSS DR12 & $0.59$ & $0.452 \pm 0.061$   & \\

54 & BOSS DR12 & $0.64$ & $0.379 \pm 0.054$   & \\

55 & SDSS DR7 & $0.1$ & $0.376\pm 0.038$  & $(\Omega_{m},\sigma_8)=(0.282,0.817)$ \\

56 & SDSS-IV & $1.52$ & $0.420 \pm 0.076$  & $(\Omega_{m},\sigma_8)=(0.26479,0.8)$ \\ 

57 & SDSS-IV & $1.52$ & $0.396 \pm 0.079$  & $(\Omega_{m},\sigma_8)=(0.31,0.8225)$ \\ 

58 & SDSS-IV & $0.978$ & $0.379 \pm 0.176$  &$(\Omega_{m},\sigma_8)=(0.31,0.8)$\\

59 & SDSS-IV & $1.23$ & $0.385 \pm 0.099$   & \\

60 & SDSS-IV & $1.526$ & $0.342 \pm 0.070$   & \\

61 & SDSS-IV & $1.944$ & $0.364 \pm 0.106$   & \\

62 & SDSS-IV & $0.72$ & $0.454 \pm 0.139$   & \\

63 & VIPERS PDR2 & $0.60$ & $0.49 \pm 0.12$  &  $(\Omega_{m},\sigma_8)=(0.31,0.8)$\\

64 & VIPERS PDR2 & $0.86$ & $0.46 \pm 0.09$  & \\

65 & BOSS DR12 voids & $0.57$ & $0.501 \pm 0.051$  &$(\Omega_{m},\sigma_8)=(0.307,0.8228)$
\\

66 & 2MTF 6dFGSv & $0.03$ & $0.404 \pm 0.0815$   &$(\Omega_{m},\sigma_8)=(0.3121,0.815)$ \\

\hline
\caption{The $f\sigma_8(z)$ data compilation from \cite{Kazantzidis:2018rnb} used in our calculations. In all cases of fiducial cosmology $\Omega_{K} = 0$. Parameter $h$ means the current Hubble parameter value in units of 100 km/s/Mpc.} 
\label{tab:data-rsd}\\
\end{longtable}

\section{THDE models and observational data analysis}

\begin{table}[H]
\centering
\begin{tabular}{|c|l|ccc|ccc|ccc|}
\hline
\textbf{Model\textbackslash{}Data}                                                       &                     & \multicolumn{3}{c|}{\textbf{SNeIa}}                                                             & \multicolumn{3}{c|}{\textbf{H}}                                                                & \multicolumn{3}{c|}{\textbf{BAO}}                                                              \\ \hline
\multirow{3}{*}{\textbf{$\Lambda$CDM}}                                                   & $\chi^2$            & \multicolumn{3}{c|}{545.093}                                                                   & \multicolumn{3}{c|}{41.641}                                                                    & \multicolumn{3}{c|}{\textbf{15.329}}                                                           \\
                                                                                         & $H_0$           & \multicolumn{3}{c|}{72.53}                                                                    & \multicolumn{3}{c|}{66.71}                                                                    & \multicolumn{3}{c|}{69.26}                                                                    \\
                                                                                         & $\Omega_{de}$ & \multicolumn{3}{c|}{0.634}                                                                     & \multicolumn{3}{c|}{0.693}                                                                     & \multicolumn{3}{c|}{0.704}                                                                     \\ \hline
\multirow{4}{*}{\textbf{\begin{tabular}[c]{@{}c@{}}THDE \\ $\gamma = 0.9$\end{tabular}}} &                     & \multicolumn{1}{c|}{\textbf{C = 0.7}} & \multicolumn{1}{c|}{\textbf{C = 1}} & \textbf{C = 1.2} & \multicolumn{1}{c|}{\textbf{C = 0.7}} & \multicolumn{1}{c|}{\textbf{C = 1}} & \textbf{C = 1.2} & \multicolumn{1}{c|}{\textbf{C = 0.7}} & \multicolumn{1}{c|}{\textbf{C = 1}} & \textbf{C = 1.2} \\ \cline{2-11} 
                                                                                         & $\chi^2$            & \multicolumn{1}{c|}{546.824}          & \multicolumn{1}{c|}{544.649}        & 543.994          & \multicolumn{1}{c|}{\textbf{40.202}}  & \multicolumn{1}{c|}{42.646}         & 43.834           & \multicolumn{1}{c|}{16.461}           & \multicolumn{1}{c|}{16.906}         & 18.352           \\
                                                                                         & $H_0$           & \multicolumn{1}{c|}{72.67}           & \multicolumn{1}{c|}{72.497}         & 72.380           & \multicolumn{1}{c|}{69.36}           & \multicolumn{1}{c|}{66.082}         & 64.82           & \multicolumn{1}{c|}{71.138}           & \multicolumn{1}{c|}{68.18}         & 66.961           \\
                                                                                         & $\Omega_{de}$ & \multicolumn{1}{c|}{0.630}            & \multicolumn{1}{c|}{0.703}          & 0.740            & \multicolumn{1}{c|}{0.733}            & \multicolumn{1}{c|}{0.738}          & 0.746            & \multicolumn{1}{c|}{0.733}            & \multicolumn{1}{c|}{0.735}          & 0.739            \\ \hline
\multirow{4}{*}{\textbf{\begin{tabular}[c]{@{}c@{}}THDE \\ $\gamma = 1$\end{tabular}}}   &                     & \multicolumn{1}{c|}{\textbf{C = 0.7}} & \multicolumn{1}{c|}{\textbf{C = 1}} & \textbf{C = 1.2} & \multicolumn{1}{c|}{\textbf{C = 0.7}} & \multicolumn{1}{c|}{\textbf{C = 1}} & \textbf{C = 1.2} & \multicolumn{1}{c|}{\textbf{C = 0.7}} & \multicolumn{1}{c|}{\textbf{C = 1}} & \textbf{C = 1.2} \\ \cline{2-11} 
                                                                                         & $\chi^2$            & \multicolumn{1}{c|}{546.728}          & \multicolumn{1}{c|}{544.577}        & 543.949          & \multicolumn{1}{c|}{40.249}           & \multicolumn{1}{c|}{42.703}         & 43.877           & \multicolumn{1}{c|}{16.410}           & \multicolumn{1}{c|}{16.845}         & 18.251           \\
                                                                                         & $H_0$           & \multicolumn{1}{c|}{72.66}           & \multicolumn{1}{c|}{72.46}         & 72.39           & \multicolumn{1}{c|}{69.31}           & \multicolumn{1}{c|}{66.023}         & 64.79           & \multicolumn{1}{c|}{71.105}           & \multicolumn{1}{c|}{68.14}         & 66.93           \\
                                                                                         & $\Omega_{de}$ & \multicolumn{1}{c|}{0.628}            & \multicolumn{1}{c|}{0.698}          & 0.737            & \multicolumn{1}{c|}{0.730}            & \multicolumn{1}{c|}{0.735}          & 0.741            & \multicolumn{1}{c|}{0.730}            & \multicolumn{1}{c|}{0.731}          & 0.734            \\ \hline
\multirow{4}{*}{\textbf{\begin{tabular}[c]{@{}c@{}}THDE \\ $\gamma = 1.1$\end{tabular}}} &                     & \multicolumn{1}{c|}{\textbf{C = 0.7}} & \multicolumn{1}{c|}{\textbf{C = 1}} & \textbf{C = 1.2} & \multicolumn{1}{c|}{\textbf{C = 0.7}} & \multicolumn{1}{c|}{\textbf{C = 1}} & \textbf{C = 1.2} & \multicolumn{1}{c|}{\textbf{C = 0.7}} & \multicolumn{1}{c|}{\textbf{C = 1}} & \textbf{C = 1.2} \\ \cline{2-11} 
                                                                                         & $\chi^2$            & \multicolumn{1}{c|}{546.641}          & \multicolumn{1}{c|}{544.510}        & 543.912          & \multicolumn{1}{c|}{40.287}           & \multicolumn{1}{c|}{42.748}         & 43.899           & \multicolumn{1}{c|}{16.362}           & \multicolumn{1}{c|}{16.757}         & 18.105           \\
                                                                                         & $H_0$           & \multicolumn{1}{c|}{72.65}           & \multicolumn{1}{c|}{72.45}         & 72.38           & \multicolumn{1}{c|}{69.23}           & \multicolumn{1}{c|}{65.93}         & 64.70           & \multicolumn{1}{c|}{71.067}           & \multicolumn{1}{c|}{68.01}         & 66.94           \\
                                                                                         & $\Omega_{de}$ & \multicolumn{1}{c|}{0.626}            & \multicolumn{1}{c|}{0.695}          & 0.732            & \multicolumn{1}{c|}{0.727}            & \multicolumn{1}{c|}{0.730}          & 0.734            & \multicolumn{1}{c|}{0.727}            & \multicolumn{1}{c|}{0.726}          & 0.729            \\ \hline
\multirow{4}{*}{\textbf{\begin{tabular}[c]{@{}c@{}}THDE\\ $\gamma = 1.2$\end{tabular}}}  &                     & \multicolumn{1}{c|}{\textbf{C = 0.7}} & \multicolumn{1}{c|}{\textbf{C = 1}} & \textbf{C = 1.2} & \multicolumn{1}{c|}{\textbf{C = 0.7}} & \multicolumn{1}{c|}{\textbf{C = 1}} & \textbf{C = 1.2} & \multicolumn{1}{c|}{\textbf{C = 0.7}} & \multicolumn{1}{c|}{\textbf{C = 1}} & \textbf{C = 1.2} \\ \cline{2-11} 
                                                                                         & $\chi^2$            & \multicolumn{1}{c|}{546.560}          & \multicolumn{1}{c|}{544.452}        & \textbf{543.884} & \multicolumn{1}{c|}{40.317}           & \multicolumn{1}{c|}{42.780}         & 43.897           & \multicolumn{1}{c|}{16.312}           & \multicolumn{1}{c|}{16.647}         & 17.913           \\
                                                                                         & $H_0$           & \multicolumn{1}{c|}{72.62}           & \multicolumn{1}{c|}{72.45}         & 72.37           & \multicolumn{1}{c|}{69.20}           & \multicolumn{1}{c|}{65.87}         & 64.66           & \multicolumn{1}{c|}{71.03}           & \multicolumn{1}{c|}{68.04}         & 66.90           \\
                                                                                         & $\Omega_{de}$ & \multicolumn{1}{c|}{0.623}            & \multicolumn{1}{c|}{0.692}          & 0.728            & \multicolumn{1}{c|}{0.725}            & \multicolumn{1}{c|}{0.725}          & 0.728            & \multicolumn{1}{c|}{0.725}            & \multicolumn{1}{c|}{0.722}          & 0.724            \\ \hline
\end{tabular}
\caption{The results for best-fit parameters $H_0$ and $\Omega_{de}$ for THDE models with some $C$ and $\gamma$ from various datasets of observations. For $BAO$ data we assume that $r_d = 147.09$ Mpc. The corresponding values of $\chi^2$ for each dataset are given. These results can be compared with standard $\Lambda$CDM model. The THDE models with better concordance in comparison with $\Lambda CDM$ model are highlighted in bold.}
\label{Tab_4}
\end{table}

\begin{figure}[b]
\centering
    \includegraphics[width=0.54\textwidth]{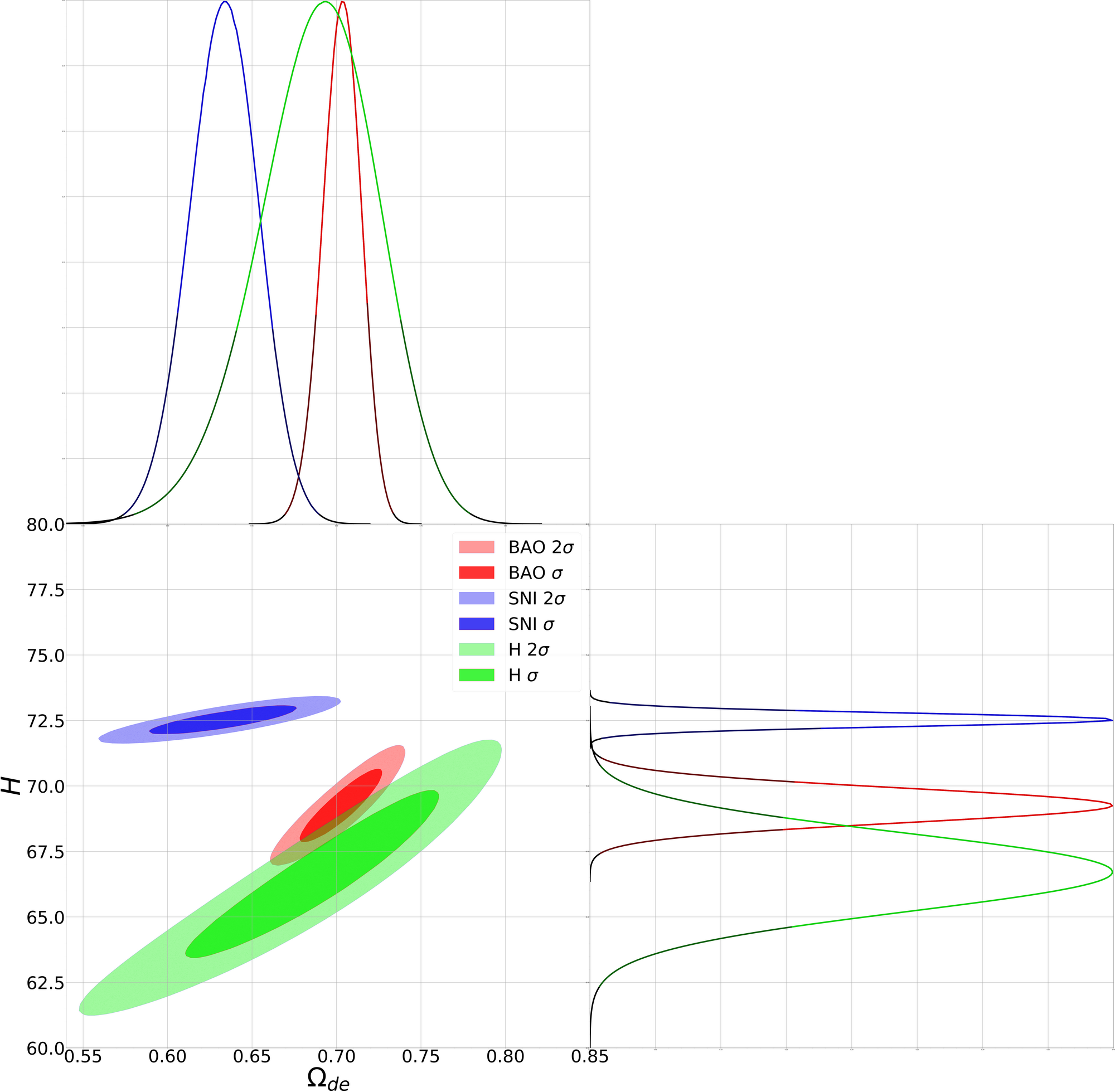}
    \caption{$1\sigma$ and $2\sigma$ areas of acceptable parameters and likelihood contours for $\Omega_{de}$ and $H_{0}$ from separate analysis of Pantheon+, BAO and $H(z)$ data for $\Lambda$CDM model. For $1\sigma$ and $2\sigma$ intervals of $\Omega_{de}$ and $H_0$ we have from  
    Pantheon+: $\Omega_{de}=0.633^{+0.029,+0.056}_{-0.027,-0.059}$, $H_0 = 72.5^{+0.38,+0.68}_{-0.30,-0.68}$; from BAO: $\Omega_{de}=0.703^{+0.015,+0.030}_{-0.015,-0.033}$, $H_0 = 69.24^{+0.91,+1.82}_{-0.91,-1.82}$; from $H(z)$: $\Omega_{de}=0.694^{+0.044,+0.085}_{-0.053,-0.114}$, $H_0 = 66.74^{+2.05,+4.02}_{-2.12,-4.39}$.} 
    \label{fig_1}
\end{figure}

\begin{figure}[H]
\centering
    \includegraphics[width=0.54\textwidth]{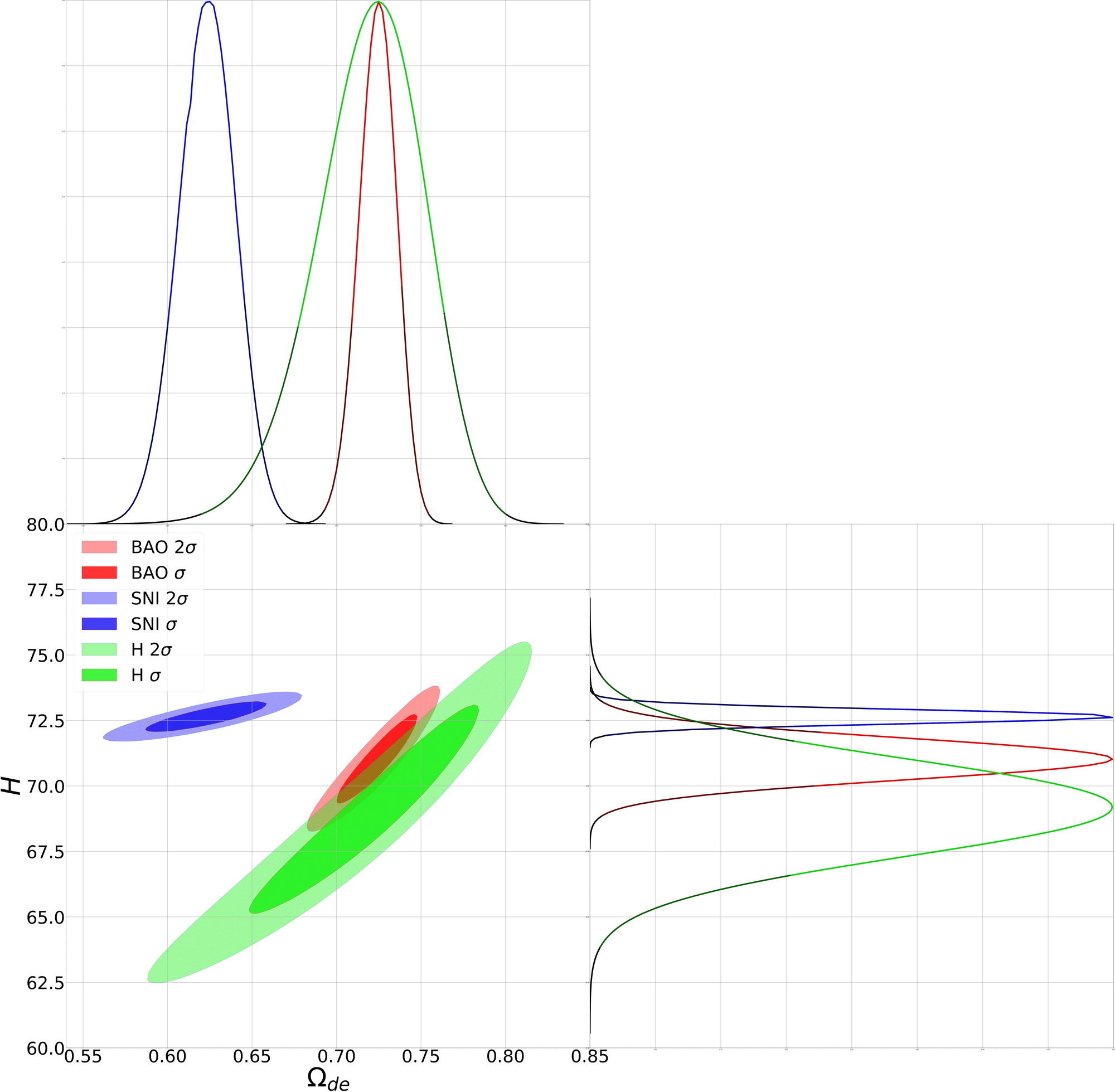}
    \caption{Similar to Fig. \ref{fig_1} but for THDE model with $C=0.7$ and $\gamma=1.2$. For $1\sigma$ and $2\sigma$ intervals of $\Omega_{de}$ and $H_0$ we have from Pantheon+: $\Omega_{de}=0.625^{+0.020,+0.043}_{-0.025,-0.050}$, $H_0 = 72.61^{+0.34,+0.80}_{-0.34,-0.68}$; from BAO: $\Omega_{de}=0.725^{+0.014,+0.030}_{-0.016,-0.032}$, $H_0 = 71.02^{+1.02,+2.16}_{-1.02,-2.16}$; from $H(z)$:  $\Omega_{de}=0.725^{+0.039,+0.075}_{-0.048,-0.105}$, $H_0 = 69.20^{+2.50,+5.00}_{-2.61,-5.34}$.} 
    \label{fig_2}
\end{figure}

\begin{figure}[H]
\centering
    \includegraphics[width=0.54\textwidth]{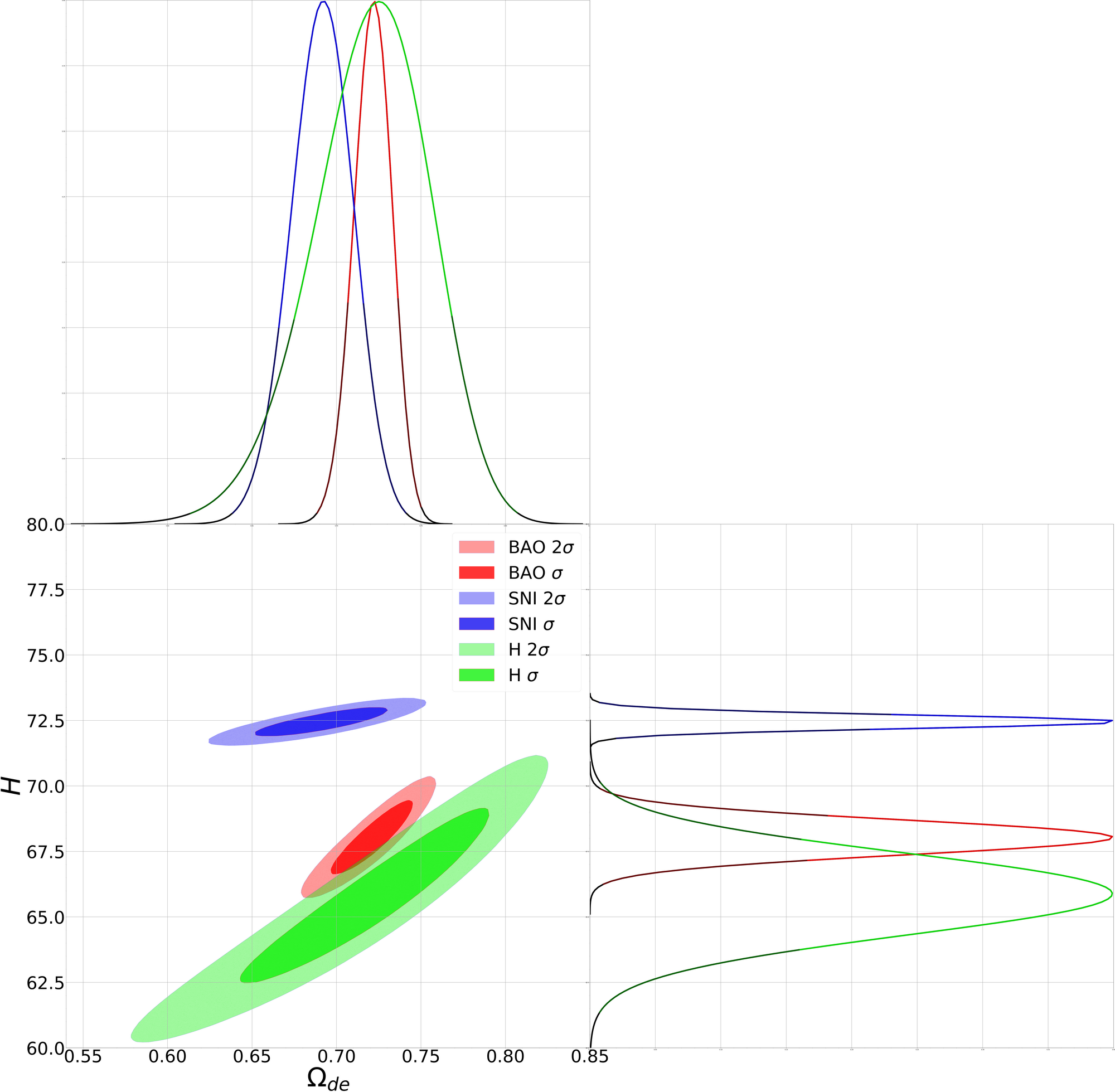}
    \caption{Similar to Fig. \ref{fig_1} but for THDE model with $C=1$ and $\gamma=1.2$. For $1\sigma$ and $2\sigma$ intervals of $\Omega_{de}$ and $H_0$ we have from Pantheon+: $\Omega_{de}=0.693^{+0.023,+0.048}_{-0.027,-0.055}$, $H_0 = 72.5^{+0.23,+0.68}_{-0.34,-0.68}$; from BAO: $\Omega_{de}=0.723^{+0.014,+0.027}_{-0.016,-0.034}$, $H_0 = 68.07^{+0.80,+1.82}_{-0.91,-1.82}$; from $H(z)$:  $\Omega_{de}=0.725^{+0.043,+0.082}_{-0.050,-0.111}$, $H_0 = 65.91^{+2.05,+4.20}_{-2.16,-4.55}$.} 
    \label{fig_3}
\end{figure}

\begin{figure}[H]
\centering
    \includegraphics[width=0.54\textwidth]{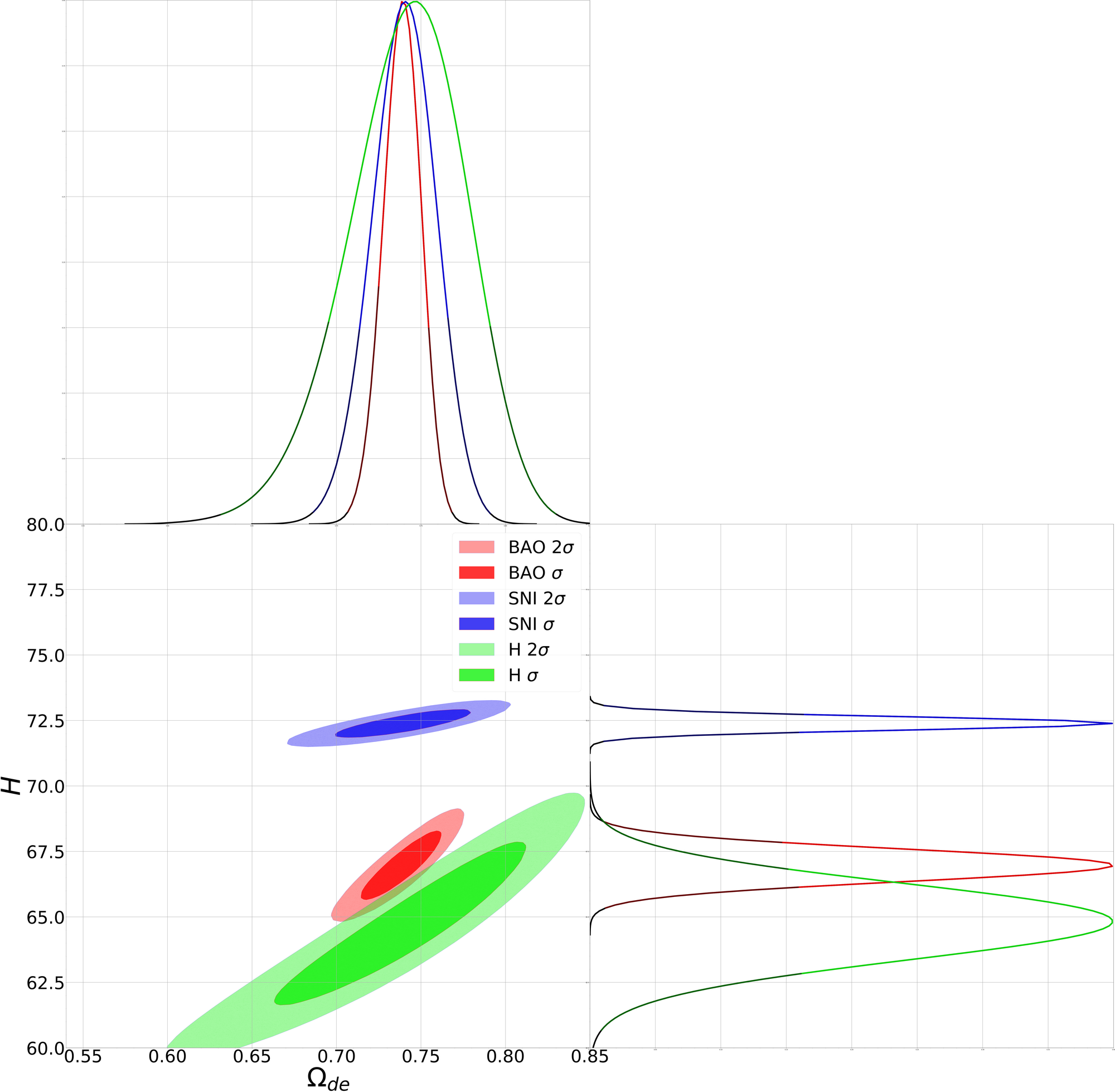}
    \caption{Similar to Fig. \ref{fig_1} but for THDE model with $C=1.2$ and $\gamma=0.9$. For $1\sigma$ and $2\sigma$ intervals of $\Omega_{de}$ and $H_0$ we have from Pantheon+: $\Omega_{de}=0.741^{+0.025,+0.050}_{-0.027,-0.055}$, $H_0 = 72.39^{+0.34,+0.68}_{-0.34,-0.68}$; from BAO: $\Omega_{de}=0.739^{+0.016,+0.030}_{-0.014,-0.032}$, $H_0 = 66.93^{+0.91,+1.70}_{-0.80,-1.59}$; from $H(z)$:  $\Omega_{de}=0.748^{+0.043,+0.082}_{-0.052,-0.116}$, $H_0 = 64.77^{+2.05,+3.98}_{-1.93,-4.09}$.} 
    \label{fig_4}
\end{figure}

We start our calculations from separate analysis of data from Pantheon+, BAO, Hubble parameter measumerements and matter density perturbations. The parameter of non-additivity as value of $C$ is fixed. Then we find point on plane $\Omega_{de} - H_{0}$ for which $\chi^2$ is minimal for given dataset. The $\Lambda$CDM model with best-fit parameters is considered as a standard for comparison with THDE models. Four values of $C$ and three values of $\gamma$ are investigated. Results for best-fit parameters are given in Table \ref{Tab_4}. SNe data favors $C>1$, the likelihood of model in fact does not depend on value of $\gamma$ (at least within the considered limits). On the contrary BAO data and Hubble parameter measurements are fitted by THDE model with better accuracy for $C<1$. 

We also depicted corresponding $1\sigma$ and $2\sigma$ areas for some $C$ and $\gamma$  on $\Omega_{de} - H_0$ plane for each dataset and likelihood contours for $\Omega_{de}$ and $H_0$ which were obtained by maximization of likelihood on remaining parameter (see Figs. \ref{fig_2}, \ref{fig_3}, \ref{fig_4}). On Fig. \ref{fig_1} one can see corresponding areas and contours for $\Lambda$CDM model. One needs to mention that the position of BAO contour on vertical axis depends on chosen $r_d$. For $r_d>149$ Mpc contour shifts down and area of overlap with contour for SNe grows. Well-known Hubble tension problem takes place: intervals of acceptable $\Omega_{de}$ from $H(z)$ data and Pantheon+ overlap but best-fit values for $H_0$ differ significantly. For THDE model the situation changes, although the problem remains. 

\begin{table}[H]
\centering
\begin{tabular}{|l|cccc|}
\hline
\multicolumn{1}{|c|}{\textbf{Par.\textbackslash{}Model}}  & \multicolumn{1}{c|}{\textbf{\begin{tabular}[c]{@{}c@{}}THDE \\ $\gamma = 0.9$\end{tabular}}} & \multicolumn{1}{c|}{\textbf{\begin{tabular}[c]{@{}c@{}}THDE \\ $\gamma = 1$\end{tabular}}} & \multicolumn{1}{c|}{\textbf{\begin{tabular}[c]{@{}c@{}}THDE \\ $\gamma = 1.1$\end{tabular}}} & \textbf{\begin{tabular}[c]{@{}c@{}}THDE \\ $\gamma = 1.2$\end{tabular}} \\ \hline
                                                                                                   & \multicolumn{4}{c|}{\textbf{C = 0.7}}                                                                                                                                                                                                                                                                                                                              \\ \hline
\textbf{$\chi^2$}                                                                          & \multicolumn{1}{c|}{32.908}                                                                 & \multicolumn{1}{c|}{32.914}                                                               & \multicolumn{1}{c|}{32.923}                                                                 & 32.932                                                   \\
\textbf{$\Omega_{de}$}                                                                  & \multicolumn{1}{c|}{0.736$^{+0.041}_{-0.041}$}                                                                   & \multicolumn{1}{c|}{0.738$^{+0.040}_{-0.045}$}                                                                 & \multicolumn{1}{c|}{0.738$^{+0.040}_{-0.045}$}                                                                   & 0.738$^{+0.040}_{-0.045}$                                                         \\ \hline

                                                                                                     & \multicolumn{4}{c|}{\textbf{C = 1}}                                                                                                                                                                                                                                                                                                                                \\ \hline
\textbf{$\chi^2$}                                                                            & \multicolumn{1}{c|}{32.430}                                                                 & \multicolumn{1}{c|}{32.436}                                                               & \multicolumn{1}{c|}{32.443}                                                                 & 32.450                                                                 \\ 
\textbf{$\Omega_{de}$}                                                                         & \multicolumn{1}{c|}{0.751$^{+0.040}_{-0.047}$}                                                                   & \multicolumn{1}{c|}{0.749$^{+0.042}_{-0.045}$}                                                                 & \multicolumn{1}{c|}{0.749$^{+0.040}_{-0.045}$}                                                                   & 0.749$^{+0.040}_{-0.047}$                                                                   \\ \hline
                                                                                                     & \multicolumn{4}{c|}{\textbf{C = 1.2}}                                                                                                                                                                                                                                                                                                                              \\ \hline
\textbf{$\chi^2$}                                                                                    & \multicolumn{1}{c|}{\textbf{32.320}}                                                                 & \multicolumn{1}{c|}{32.327}                                                               & \multicolumn{1}{c|}{32.334}                                                                 & 32.343                                                                 \\ 
\textbf{$\Omega_{de}$}                                                                         & \multicolumn{1}{c|}{\textbf{0.756}$^{+0.040}_{-0.047}$}                                                                   & \multicolumn{1}{c|}{0.755$^{+0.042}_{-0.045}$}                                                                 & \multicolumn{1}{c|}{0.755$^{+0.040}_{-0.047}$}                                                                   & 0.753$^{+0.042}_{-0.045}$                                                                   \\ \hline
\end{tabular}
\caption{The results for best-fit values of $\Omega_{de}$ with $1\sigma$ intervals from analysis of $f\sigma_8$ data for THDE models with various $C$ and $\gamma$. The values of $\chi^2_{}$ are very close to standard cosmological model ($\chi^{2(\Lambda CDM)} = 32.639, \Omega_{de} = 0.745^{+0.040}_{-0.045}$), for $C=1$ and $C=1.2$ $\chi^2<\chi^{2(\Lambda CDM)}$.}
\label{Tab_5}
\end{table}

\begin{figure}
\centering
    \includegraphics[width=0.8\textwidth]{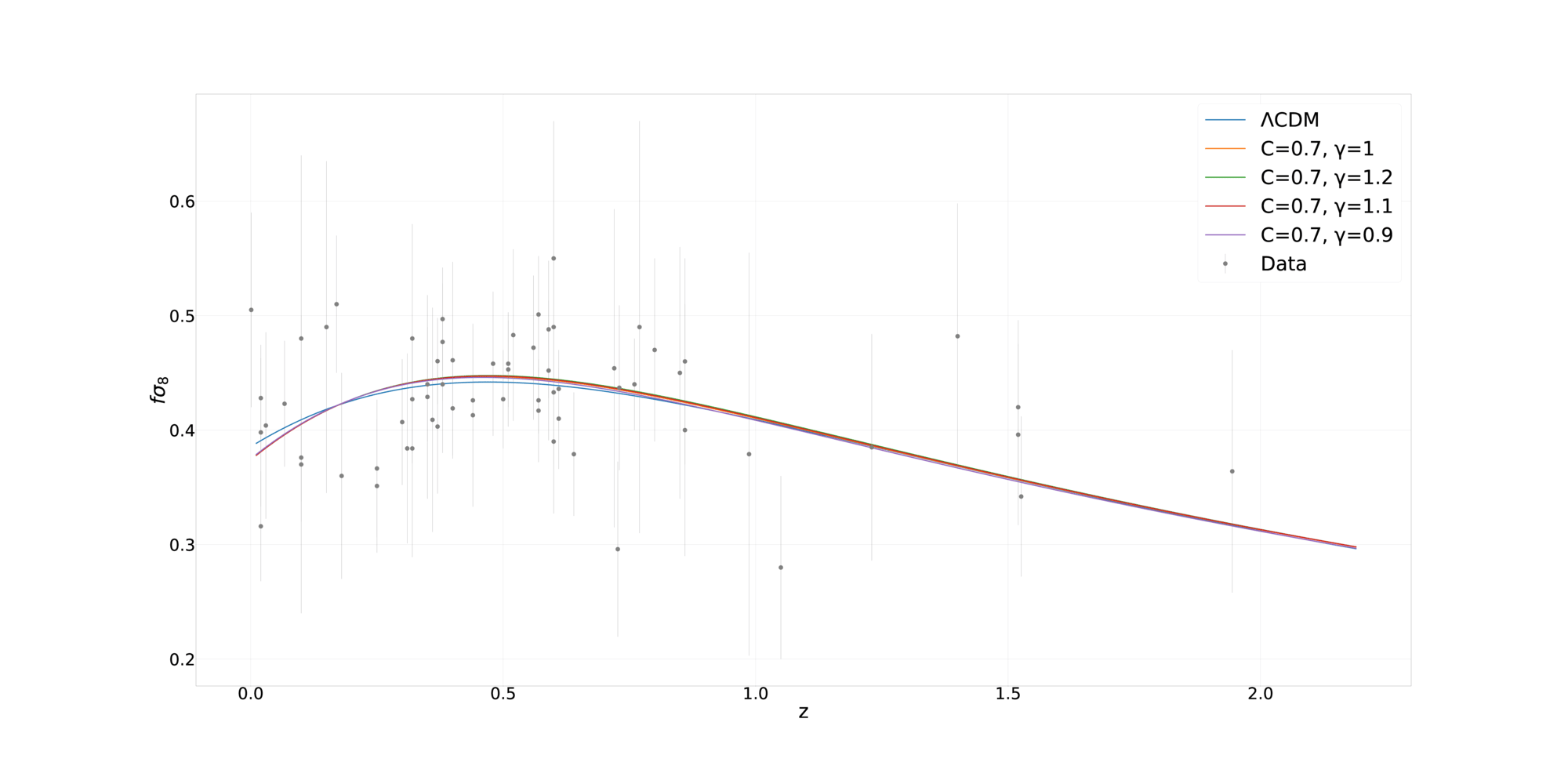}\\
    \includegraphics[width=0.8\textwidth]{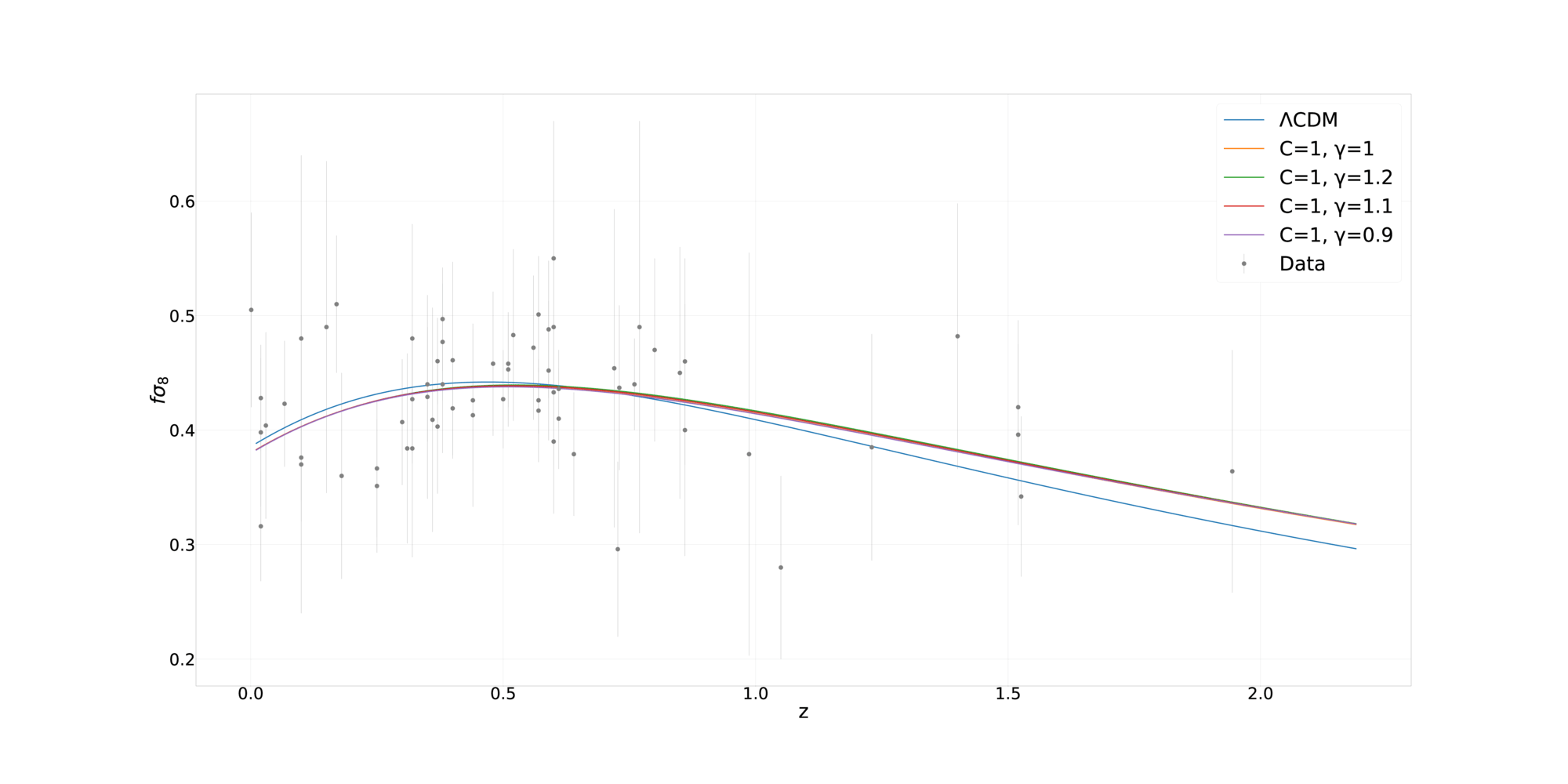}\\
    \includegraphics[width=0.8\textwidth]{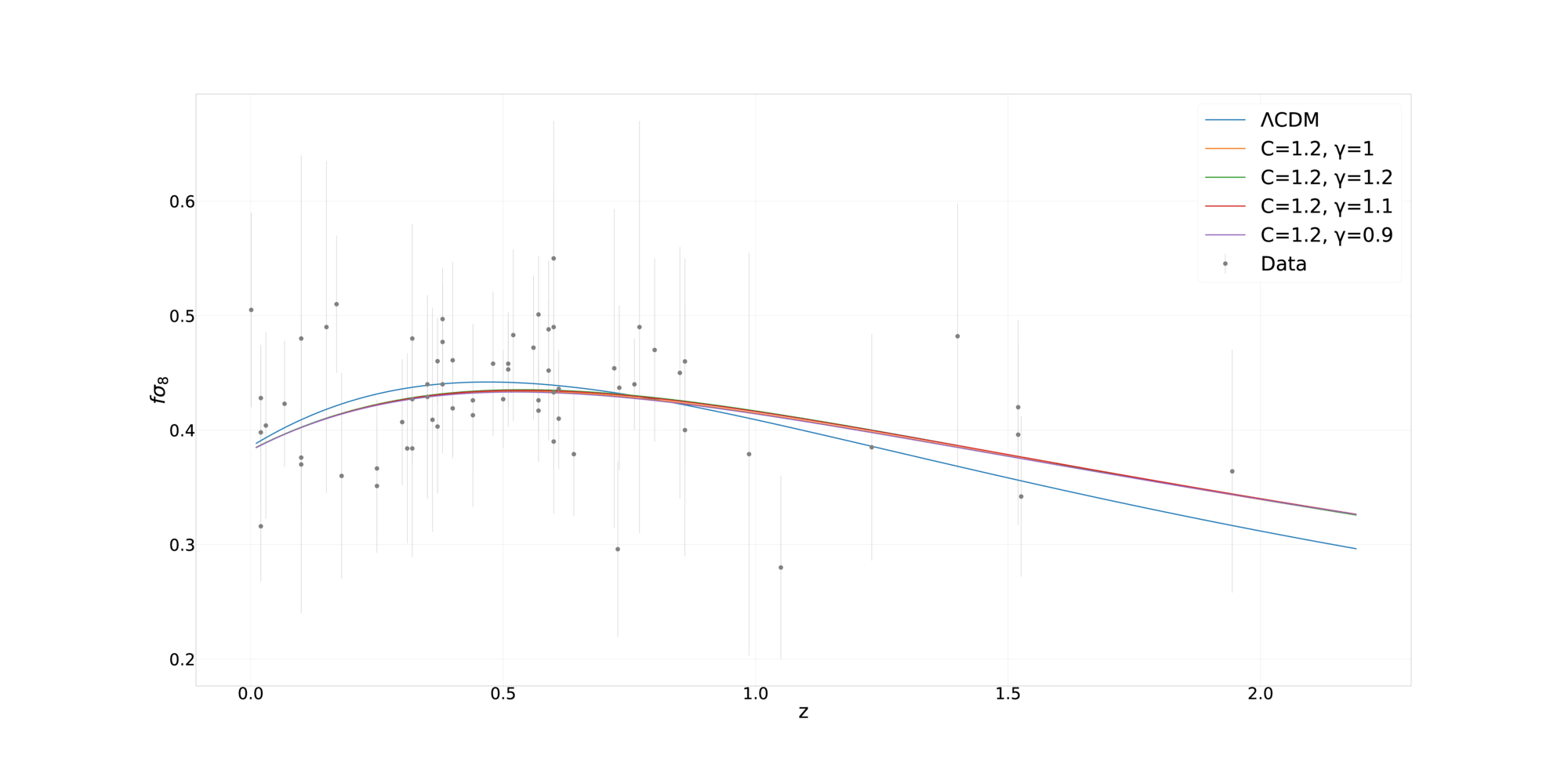}
    \caption{Dependence of $f\sigma_8$ on redshift for $C=0.7$ (upper panel), $C=1$ (middle panel), $C=1.2$ (down panel) and some values of $\gamma$ in comparison with $\Lambda$CDM model. The parameters $\Omega_{de}$ and $\sigma_8$ are fixed for maximal likelihood function. One can see that there is no significant difference between models with various $\gamma$ for given $C$. For $C\geq 1$ from $z>1$ deviations from $\Lambda$CDM model grow up. } 
    \label{fig_5}
\end{figure}

The acceptable interval for $\Omega_{de}$ from matter density perturbations data is close to that from BAO, $H(z)$ data for all considered $C$ and $\gamma$ and from Pantheon+ if $C>1$ (Table \ref{Tab_5}). As in the case of SNe analysis of matter perturbations data $f\sigma_8$ favors $C>1$. The evolution of perturbations for various $C$ and $\gamma$ in THDE model is depicted on Fig. \ref{fig_5}. The dependence $f\sigma_{8}(z)$ is very close to that in standard cosmology, some deviations take place only for $z>\sim 1$.

\begin{table}
\centering
\begin{tabular}{|l|cccc|}
\hline
\multicolumn{1}{|c|}{\textbf{Par.\textbackslash{}Model}}  & \multicolumn{1}{c|}{\textbf{\begin{tabular}[c]{@{}c@{}}THDE \\ $\gamma = 0.9$\end{tabular}}} & \multicolumn{1}{c|}{\textbf{\begin{tabular}[c]{@{}c@{}}THDE \\ $\gamma = 1$\end{tabular}}} & \multicolumn{1}{c|}{\textbf{\begin{tabular}[c]{@{}c@{}}THDE \\ $\gamma = 1.1$\end{tabular}}} & \textbf{\begin{tabular}[c]{@{}c@{}}THDE \\ $\gamma = 1.2$\end{tabular}} \\ \hline
                                                                                                 & \multicolumn{4}{c|}{\textbf{C = 0.7}}                                                                                                                                                                                                                                                                                                                              \\ \hline
\textbf{$\chi^2$}                                            & \multicolumn{1}{c|}{674.833}                   & \multicolumn{1}{c|}{674.695}                  & \multicolumn{1}{c|}{674.549}                & \textbf{674.418}                                                        \\ 
\textbf{$H_0$}                                                                          & \multicolumn{1}{c|}{74.21}             & \multicolumn{1}{c|}{74.18}          & \multicolumn{1}{c|}{74.19}                                                                  & \textbf{74.17}       \\ 
\textbf{$\Omega_{de}$}                                                                  & \multicolumn{1}{c|}{0.762}                                                                   & \multicolumn{1}{c|}{0.760}                                                                 & \multicolumn{1}{c|}{0.758}                                                                   & \textbf{0.756}                                                          \\ \hline
                                                                                                    & \multicolumn{4}{c|}{\textbf{C = 1}}                                                                                                                                                                                                                                                                                                                                \\ \hline
\textbf{$\chi^2$}                                                                                    & \multicolumn{1}{c|}{688.519}                                                                 & \multicolumn{1}{c|}{688.911}                                                               & \multicolumn{1}{c|}{689.125}                                                                 & 689.188                                                                 \\ 
\textbf{$H_0$}                                                                                   & \multicolumn{1}{c|}{73.10}                                                                  & \multicolumn{1}{c|}{73.09}                                                                & \multicolumn{1}{c|}{73.06}                                                                  & 73.03                                                                  \\ 
\textbf{$\Omega_{de}$}                                                                        & \multicolumn{1}{c|}{0.797}                                                                   & \multicolumn{1}{c|}{0.794}                                                                 & \multicolumn{1}{c|}{0.791}                                                                   & 0.787                                                                   \\ \hline
                                                                                                    & \multicolumn{4}{c|}{\textbf{C = 1.2}}                                                                                                                                                                                                                                                                                                                              \\ \hline
\textbf{$\chi^2$}                                                                                    & \multicolumn{1}{c|}{709.707}                                                                 & \multicolumn{1}{c|}{709.997}                                                               & \multicolumn{1}{c|}{709.923}                                                                 & 709.494                                                                 \\ 
\textbf{$H_0$}                                                                                   & \multicolumn{1}{c|}{72.54}                                                                  & \multicolumn{1}{c|}{72.55}                                                                & \multicolumn{1}{c|}{72.52}                                                                  & 72.54                                                                  \\ 
\textbf{$\Omega_{de}$}                                                                         & \multicolumn{1}{c|}{0.815}                                                                   & \multicolumn{1}{c|}{0.811}                                                                 & \multicolumn{1}{c|}{0.806}                                                                   & 0.802                                                                   \\ \hline
\end{tabular}
\caption{The results for best-fit values of  $\Omega_{de}$ and $H_0$ from combined Pantheon+BAO+$H(z)$ analysis of various THDE models. Corresponding values of $\chi^2$ are given. For comparison best-fit values  for $\Lambda$CDM model are $H_0 = 73.31, \Omega_{de} = 0.757$ ($\chi^2 = 676.755$).}
\label{Tab_6}
\end{table}

\begin{figure}[H]
\centering
    \includegraphics[width=0.65\textwidth]{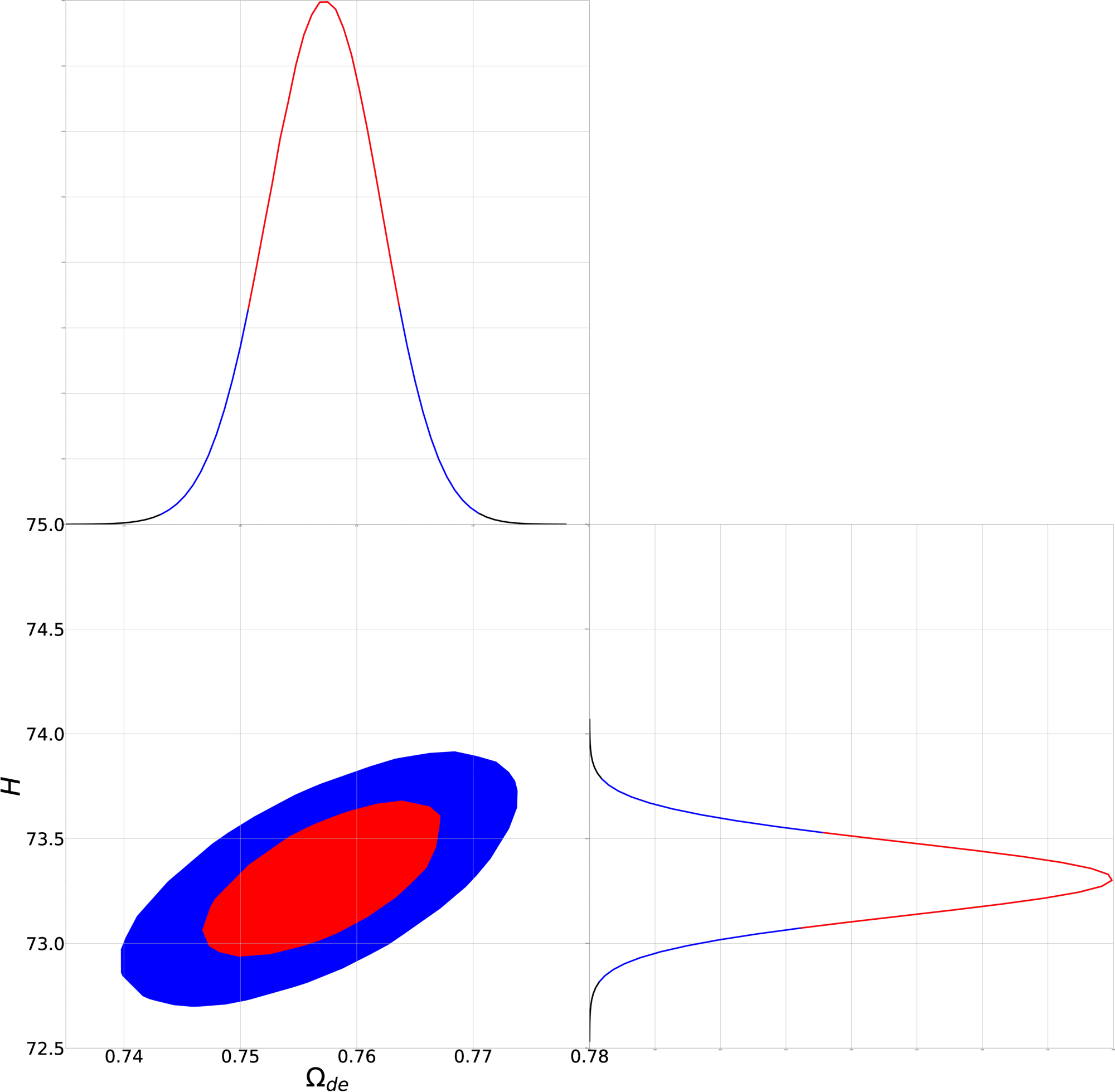}
    \caption{$1\sigma$ and $2\sigma$ areas of acceptable parameters ($\Omega_{de}, H_{0}$) and likelihood contours from combined analysis of Pantheon+BAO+$H(z)$ data for $\Lambda$CDM model. The optimal values of parameters with $1\sigma$ and $2\sigma$ errors are $\Omega_{de}=0.758^{+0.006,+0.013}_{-0.007,-0.014}$, $H_0 = 73.30^{+0.23,+0.48}_{-0.23,-0.48}$.} 
    \label{fig_6}
\end{figure}

\begin{figure}[H]
\centering
    \includegraphics[width=0.5\textwidth]{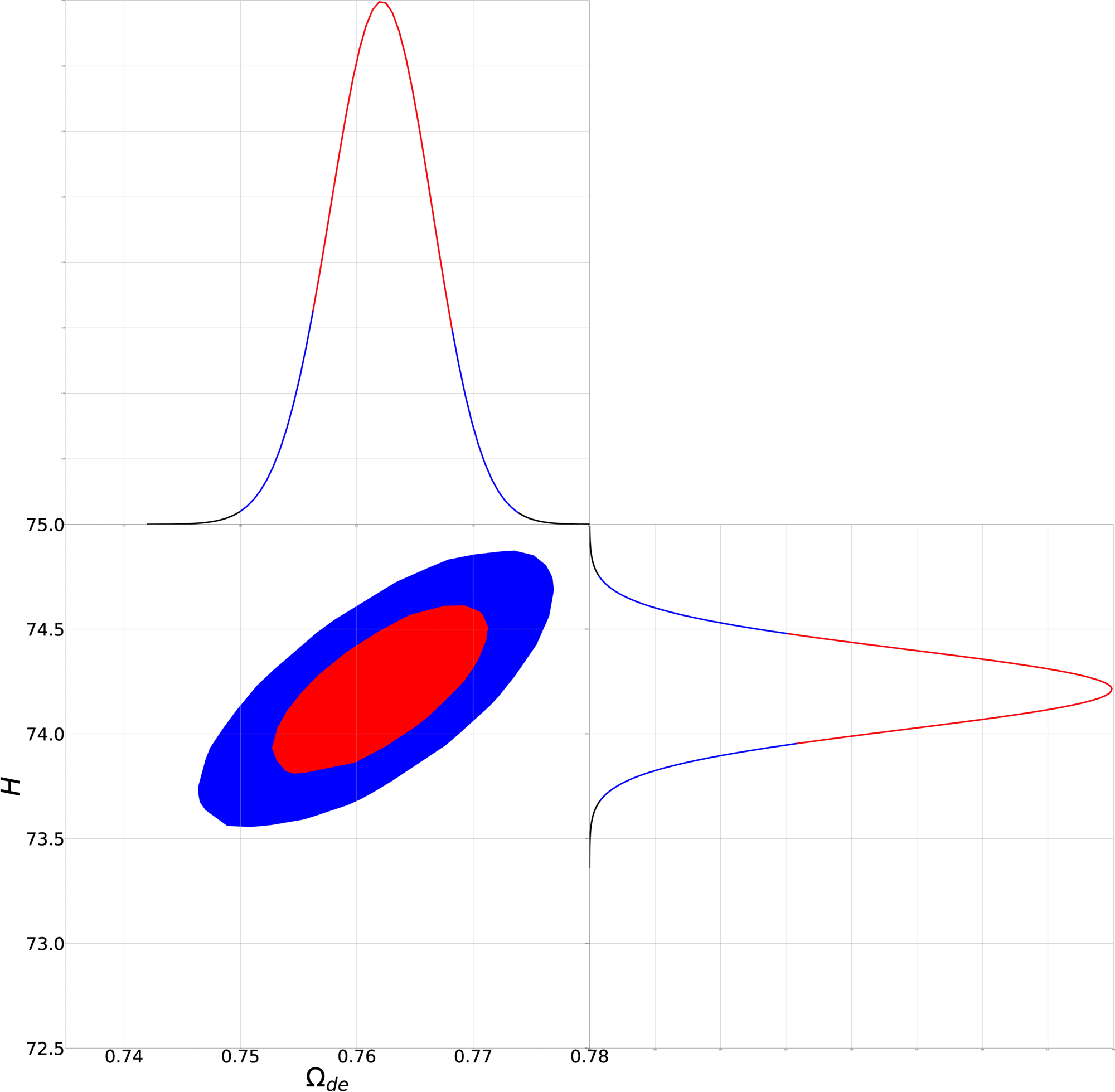}\includegraphics[width=0.5\textwidth]{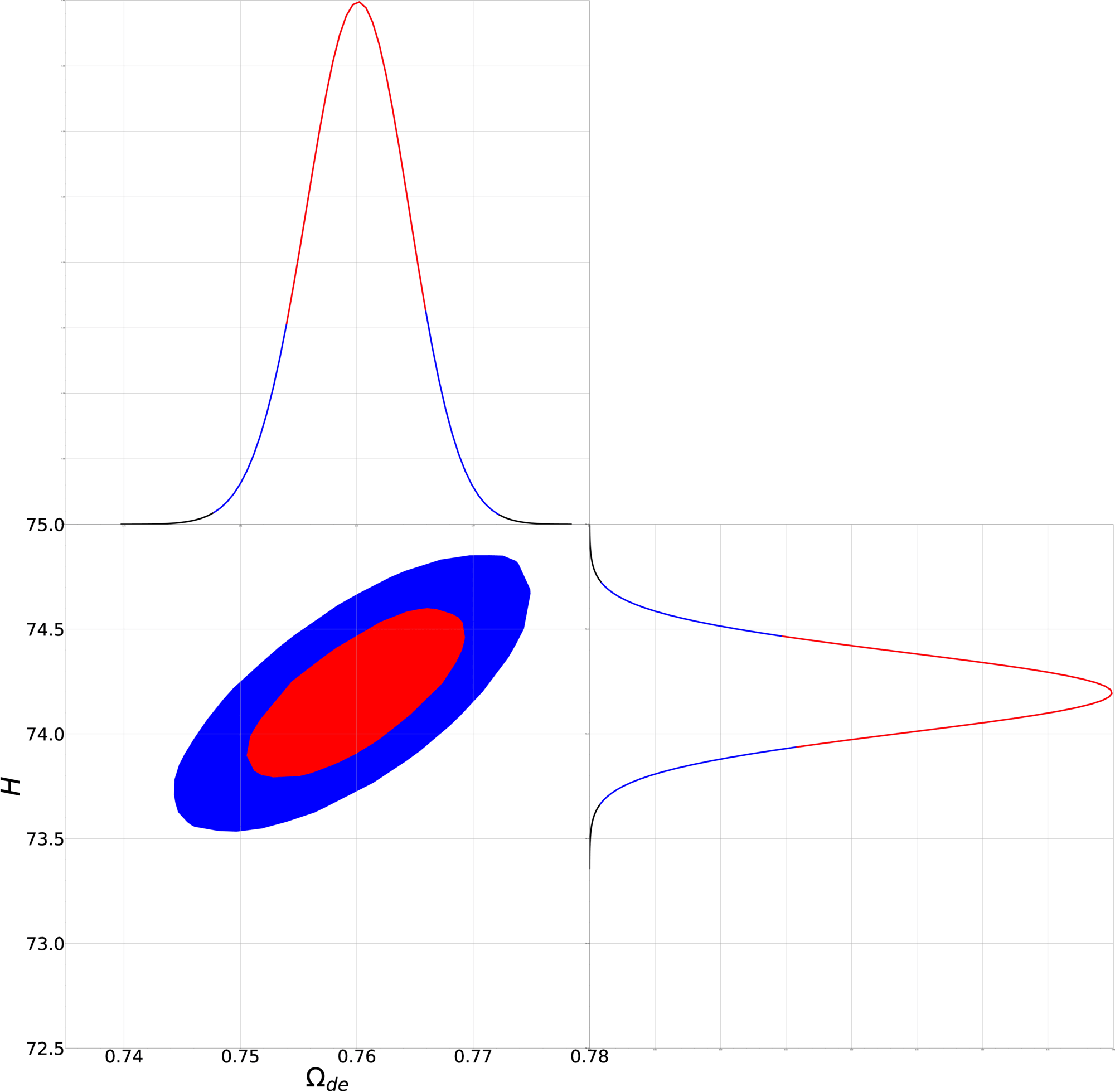}\\
     \includegraphics[width=0.5\textwidth]{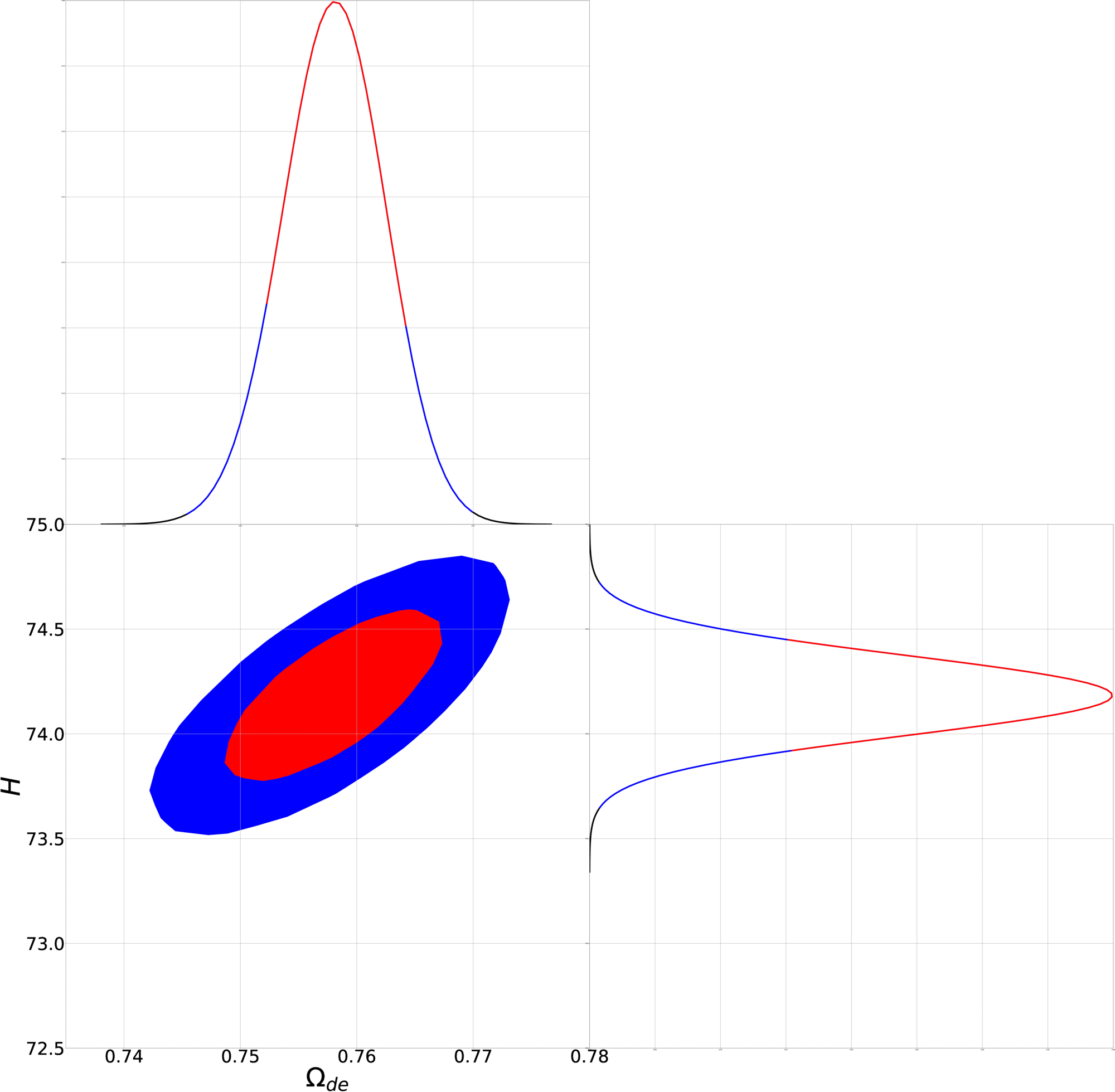}\includegraphics[width=0.5\textwidth]{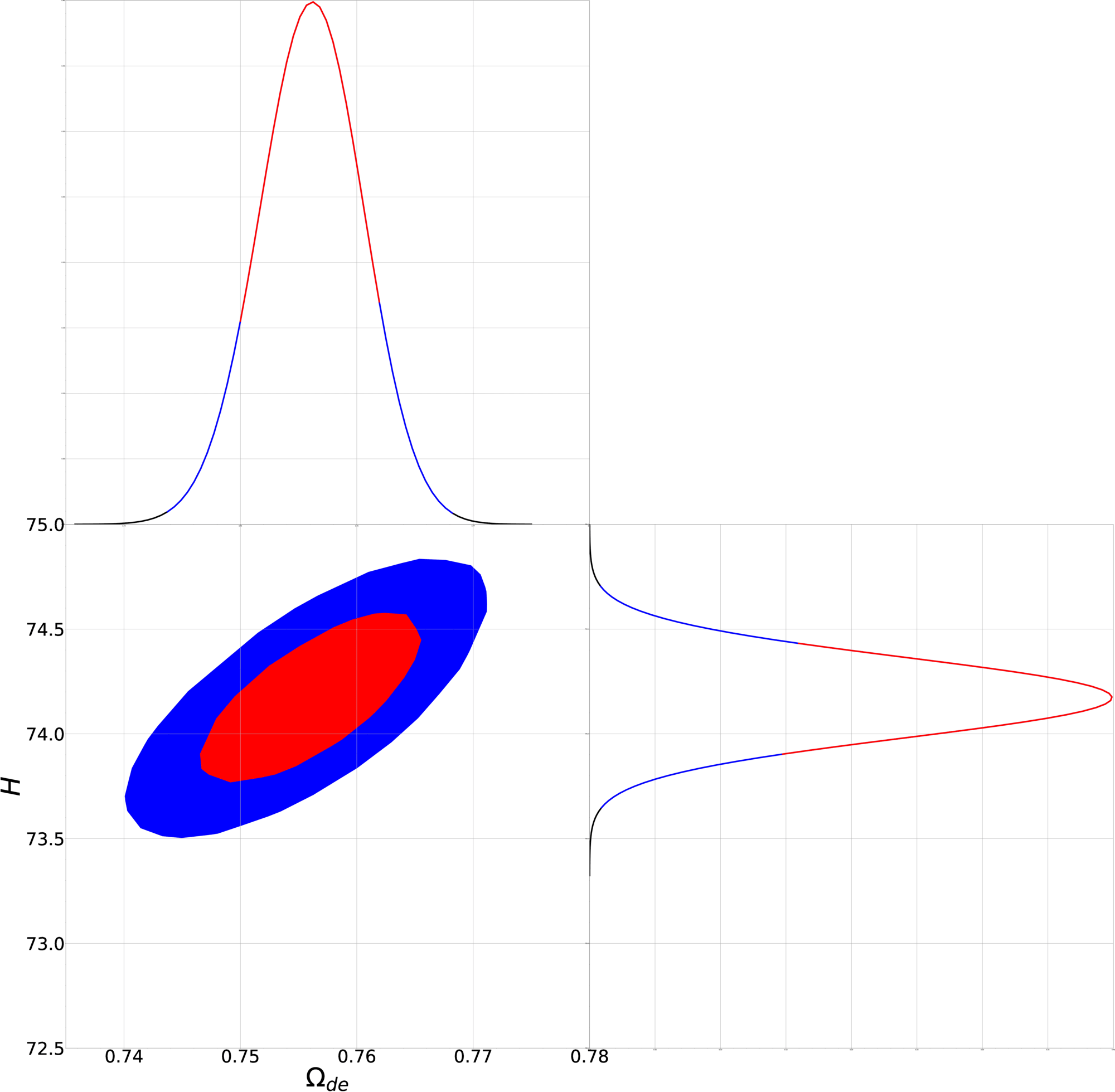}
    \caption{Similar to Fig. \ref{fig_6} but for holographic dark energy with $C=0.7$ and $\gamma=0.9$ (left upper panel), $\gamma=1.0$ (right upper panel), $\gamma=1.1$ (left down panel), $\gamma=1.2$ (right down panel).  The best-fit values of $\Omega_{de}$ and $H_0$ are for $\gamma=0.9$:  $\Omega_{de}=0.762^{+0.006,+0.012}_{-0.006,-0.012}$, $H_0 = 74.22^{+0.26,+0.53}_{-0.26,-0.53}$; $\gamma=1.0$: $\Omega_{de}=0.760^{+0.006,+0.012}_{-0.006,-0.013}$, $H_0 = 74.19^{+0.27,+0.53}_{-0.26,-0.53}$; $\gamma=1.1$: $\Omega_{de}=0.758^{+0.006,+0.012}_{-0.006,-0.012}$, $H_0 = 74.18^{+0.27,+0.55}_{-0.26,-0.53}$; $\gamma=1.2$: $\Omega_{de}=0.756^{+0.006,+0.012}_{-0.006,-0.012}$, $H_0 = 74.18^{+0.26,+0.53}_{-0.27,-0.53}$.} 
    \label{fig_7}
\end{figure}

The results of joint analysis of observational data (Pantheon+BAO+$H(z)$) are presented in table \ref{Tab_6}. Consideration of all data allows us to conclude that value $C=0.7$ is preferred while in the case of $C=1$ and $C=1.2$ the likelihood of THDE model is significantly less. Also, there is no substantial dependence from parameter $\gamma$. It can also be noted that the acceptable range of $\Omega_{de}$ from $f\sigma_8$ data overlaps with results from combined remaining data.

The $1\sigma$ and $2\sigma$ areas of acceptable parameters are given on Fig. \ref{fig_6} for $\Lambda$CDM model and on Fig. \ref{fig_7} for $C=0.7$, $\gamma=0.9$, $1.0$, $1.1$, $1.2$. The best-fit value for $H_0$ in THDE model is $74.2$ km/s/Mpc, which is slightly larger in comparison to $\Lambda$CDM model. 

\begin{table}
\centering
\begin{tabular}{|l|cccc|}
\hline
\multicolumn{1}{|c|}{\textbf{Par.\textbackslash{}Model}}  & \multicolumn{1}{c|}{\textbf{\begin{tabular}[c]{@{}c@{}}THDE \\ $\gamma = 0.9$\end{tabular}}} & \multicolumn{1}{c|}{\textbf{\begin{tabular}[c]{@{}c@{}}THDE \\ $\gamma = 1$\end{tabular}}} & \multicolumn{1}{c|}{\textbf{\begin{tabular}[c]{@{}c@{}}THDE \\ $\gamma = 1.1$\end{tabular}}} & \textbf{\begin{tabular}[c]{@{}c@{}}THDE\\ $\gamma = 1.2$\end{tabular}} \\ \hline
                                                                                                           & \multicolumn{4}{c|}{\textbf{C = 0.7, $\beta = 0.1$,  $\alpha = 0.1$}}                                                                                                                                                                                                                                                                                             \\ \hline
$\chi^2$                                                                      & \multicolumn{1}{c|}{677.169}                                                                 & \multicolumn{1}{c|}{676.886}                                                               & \multicolumn{1}{c|}{676.646}                                                                 & 676.494                                                                \\ \
$H_0$                                                                     & \multicolumn{1}{c|}{74.564}                                                                  & \multicolumn{1}{c|}{74.530}                                                                & \multicolumn{1}{c|}{74.478}                                                                  & 74.484                                                                 \\ 
$\Omega_{de}$                                                            & \multicolumn{1}{c|}{0.805}                                                                   & \multicolumn{1}{c|}{0.803}                                                                 & \multicolumn{1}{c|}{0.800}                                                                   & 0.798                                                                  \\ \hline
\multicolumn{1}{|c|}{\textbf{}}                                                                            & \multicolumn{4}{c|}{\textbf{C = 0.7, $\beta = -0.1$,  $\alpha = 0.1$}}                                                                                                                                                                                                                                                                                            \\ \hline
$\chi^2$                                                                                                  & \multicolumn{1}{c|}{674.723}                                                                 & \multicolumn{1}{c|}{674.524}                                                               & \multicolumn{1}{c|}{674.324}                                                                 & \textbf{674.191}                                                       \\ 
$H_0$                                                                                                  & \multicolumn{1}{c|}{74.332}                                                                  & \multicolumn{1}{c|}{74.319}                                                                & \multicolumn{1}{c|}{74.325}                                                                  & \textbf{74.317}                                                        \\ 
$\Omega_{de}$                                                                                       & \multicolumn{1}{c|}{0.758}                                                                   & \multicolumn{1}{c|}{0.758}                                                                 & \multicolumn{1}{c|}{0.756}                                                                   & \textbf{0.754}                                                         \\ \hline
\multicolumn{1}{|c|}{\textbf{}}                                                                           & \multicolumn{4}{c|}{\textbf{C = 0.7, $\beta = 0.1$,  $\alpha = -0.1$}}                                                                                                                                                                                                                                                                                            \\ \hline
$\chi^2$                                                                                                  & \multicolumn{1}{c|}{677.045}                                                                 & \multicolumn{1}{c|}{677.048}                                                               & \multicolumn{1}{c|}{676.999}                                                                 & 676.943                                                                \\ 
$H_0$                                                                                                 & \multicolumn{1}{c|}{74.040}                                                                  & \multicolumn{1}{c|}{74.027}                                                                & \multicolumn{1}{c|}{74.010}                                                                  & 74.002                                                                 \\ 
$\Omega_{de}$                                                                                       & \multicolumn{1}{c|}{0.764}                                                                   & \multicolumn{1}{c|}{0.763}                                                                 & \multicolumn{1}{c|}{0.759}                                                                   & 0.757                                                                  \\ \hline
\multicolumn{1}{|c|}{\textbf{}}                                                                          & \multicolumn{4}{c|}{\textbf{C = 0.7, $\beta = -0.1$,  $\alpha = -0.1$}}                                                                                                                                                                                                                                                                                           \\ \hline
$\chi^2$                                                                                                  & \multicolumn{1}{c|}{678.049}                                                                 & \multicolumn{1}{c|}{678.004}                                                               & \multicolumn{1}{c|}{677.939}                                                                 & 677.818                                                                \\ 
$H_0$                                                                                                 & \multicolumn{1}{c|}{73.932}                                                                  & \multicolumn{1}{c|}{73.881}                                                                & \multicolumn{1}{c|}{73.898}                                                                  & 73.861                                                                 \\ 
$\Omega_{de}$                                                                                       & \multicolumn{1}{c|}{0.719}                                                                   & \multicolumn{1}{c|}{0.716}                                                                 & \multicolumn{1}{c|}{0.715}                                                                   & 0.712                                                                  \\ \hline
\multicolumn{1}{|c|}{\textbf{}}                                                                           & \multicolumn{4}{c|}{\textbf{C = 1, $\beta= 0.1$,  $\alpha = 0.1$}}                                                                                                                                                                                                                                                                                               \\ \hline
$\chi^2$                                                                                                   & \multicolumn{1}{c|}{677.789}                                                                 & \multicolumn{1}{c|}{678.081}                                                               & \multicolumn{1}{c|}{678.276}                                                                 & 678.365                                                                \\ 
$H_0$                                                                                                 & \multicolumn{1}{c|}{73.366}                                                                  & \multicolumn{1}{c|}{73.326}                                                                & \multicolumn{1}{c|}{73.279}                                                                  & 73.272                                                                 \\ 
$\Omega_{de}$                                                                                        & \multicolumn{1}{c|}{0.844}                                                                   & \multicolumn{1}{c|}{0.841}                                                                 & \multicolumn{1}{c|}{0.838}                                                                   & 0.835                                                                  \\ \hline
\multicolumn{1}{|c|}{\textbf{}}                                                                            & \multicolumn{4}{c|}{\textbf{C = 1, $\beta = -0.1$,  $\alpha = 0.1$}}                                                                                                                                                                                                                                                                                              \\ \hline
$\chi^2$                                                                                                   & \multicolumn{1}{c|}{682.331}                                                                 & \multicolumn{1}{c|}{682.393}                                                               & \multicolumn{1}{c|}{682.342}                                                                 & 682.166                                                                \\ 
$H_0$                                                                                                  & \multicolumn{1}{c|}{73.162}                                                                  & \multicolumn{1}{c|}{73.153}                                                                & \multicolumn{1}{c|}{73.124}                                                                  & 73.144                                                                 \\ 
$\Omega_{de}$                                                                                        & \multicolumn{1}{c|}{0.787}                                                                   & \multicolumn{1}{c|}{0.784}                                                                 & \multicolumn{1}{c|}{0.781}                                                                   & 0.779                                                                  \\ \hline
\multicolumn{1}{|c|}{\textbf{}}                                                                            & \multicolumn{4}{c|}{\textbf{C = 1, $\beta = 0.1$,  $\alpha = -0.1$}}                                                                                                                                                                                                                                                                                              \\ \hline
$\chi^2$                                                                                                   & \multicolumn{1}{c|}{695.726}                                                                 & \multicolumn{1}{c|}{696.513}                                                               & \multicolumn{1}{c|}{697.125}                                                                 & 697.530                                                                \\ 
$H_0$                                                                                                  & \multicolumn{1}{c|}{72.965}                                                                  & \multicolumn{1}{c|}{72.958}                                                                & \multicolumn{1}{c|}{72.950}                                                                  & 72.902                                                                 \\ 
$\Omega_{de}$                                                                                        & \multicolumn{1}{c|}{0.807}                                                                   & \multicolumn{1}{c|}{0.803}                                                                 & \multicolumn{1}{c|}{0.799}                                                                   & 0.796                                                                  \\ \hline
\multicolumn{1}{|c|}{\textbf{}}                                                                            & \multicolumn{4}{c|}{\textbf{C = 1, $\beta = -0.1$,  $\alpha = -0.1$}}                                                                                                                                                                                                                                                                                             \\ \hline
$\chi^2$                                                                                                   & \multicolumn{1}{c|}{702.990}                                                                 & \multicolumn{1}{c|}{703.380}                                                               & \multicolumn{1}{c|}{703.535}                                                                 & 703.504                                                                \\ 
$H_0$                                                                                                  & \multicolumn{1}{c|}{72.856}                                                                  & \multicolumn{1}{c|}{72.845}                                                                & \multicolumn{1}{c|}{72.806}                                                                  & 72.762                                                                 \\ 
$\Omega_{de}$                                                                                        & \multicolumn{1}{c|}{0.748}                                                                   & \multicolumn{1}{c|}{0.746}                                                                 & \multicolumn{1}{c|}{0.742}                                                                   & 0.738                                                                  \\ \hline
\end{tabular}
\caption{The results for best-fit values $\Omega_{de}$ and $H_0$ from combined Pantheon+BAO+$H(z)$ data analysis for THDE models with various $C$ and $\gamma$ in presence of interaction between matter and dark component in form $Q=H(\alpha\rho_m + \beta\rho_{de})$. Thus, the results for $C=1$ are close to $\Lambda$CDM if $\alpha, \beta>0$, for other cases concordance with observational data worsens. For $C=0.7$ the model describes data with better accuracy in comparison to $\Lambda$CDM model for $\beta<0$, $\alpha>0$.}
\label{Tab_7}
\end{table}

\begin{figure}
\centering
    \includegraphics[width=0.6\textwidth]{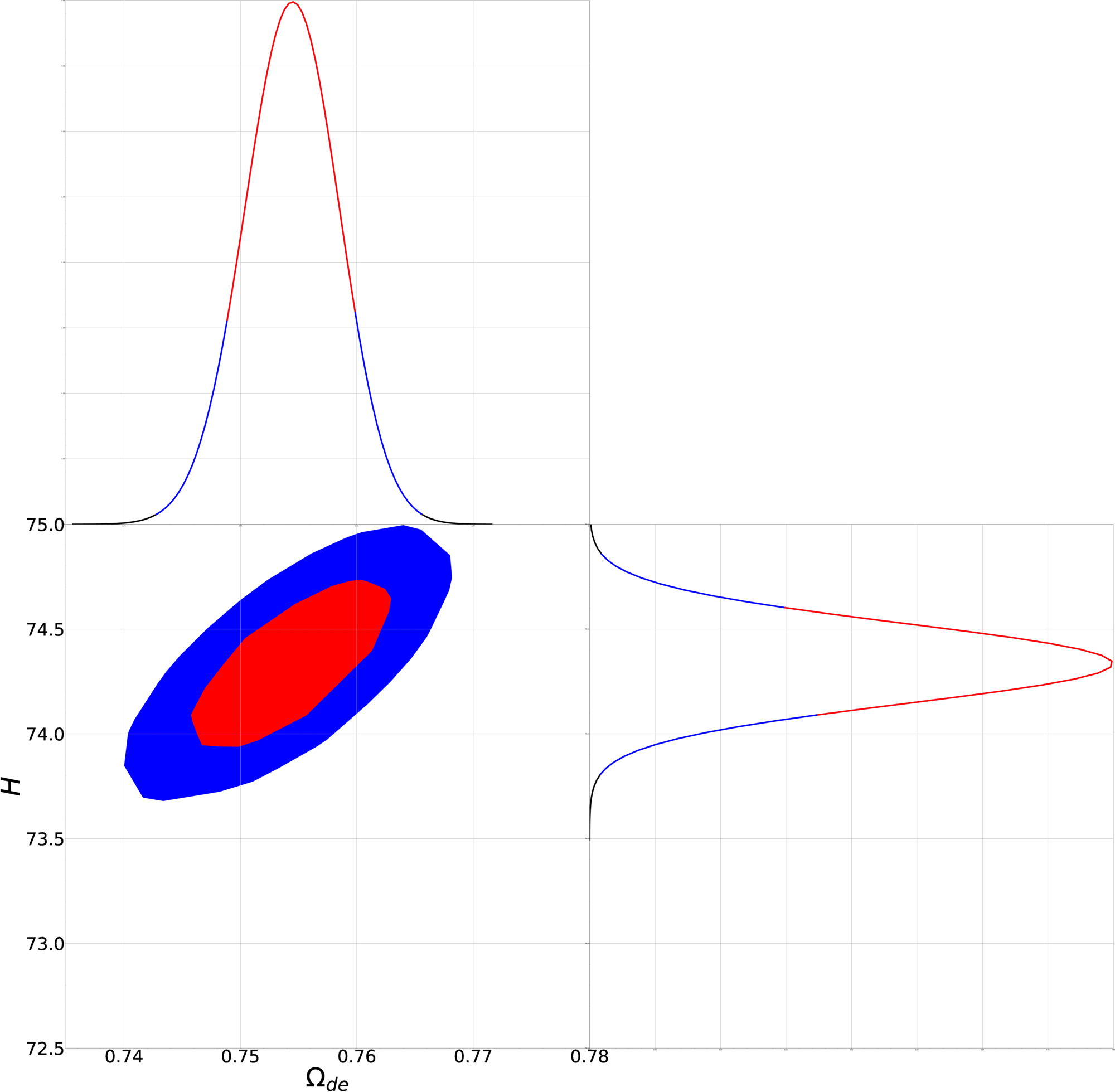}
    \caption{$1\sigma$ and $2\sigma$ areas of allowable parameters on $\Omega_{de}-H_{0}$ and likelihood contours from combined analysis of Pantheon+BAO+$H(z)$ data for THDE model in presence of interaction. The case for which concordance with observational data is better than for $\Lambda$CDM model is considered ($C = 0.7, \gamma = 1.2, \beta = -0.1, \alpha = 0.1$).} 
    \label{fig_8}
\end{figure}

\begin{table}
\centering
\begin{tabular}{|l|cccc|}
\hline
\multicolumn{1}{|c|}{\textbf{Par.\textbackslash{}Model}}  & \multicolumn{1}{c|}{\textbf{\begin{tabular}[c]{@{}c@{}}THDE \\ $\gamma = 0.9$\end{tabular}}} & \multicolumn{1}{c|}{\textbf{\begin{tabular}[c]{@{}c@{}}THDE \\ $\gamma = 1$\end{tabular}}} & \multicolumn{1}{c|}{\textbf{\begin{tabular}[c]{@{}c@{}}THDE \\ $\gamma = 1.1$\end{tabular}}} & \textbf{\begin{tabular}[c]{@{}c@{}}THDE\\ $\gamma = 1.2$\end{tabular}} \\ \hline
                                                                                                           & \multicolumn{4}{c|}{\textbf{C = 0.7, $\beta= 0.1$,  $\alpha= 0.1$}}                                                                                                                                                                                                                                                                                               \\ \hline
$\chi^2$                                                                       & \multicolumn{1}{c|}{33.135}                                                                  & \multicolumn{1}{c|}{33.144}                                                                & \multicolumn{1}{c|}{33.152}                                                                  & 33.162                                                                 \\ 
$\Omega_{de}$                                                           & \multicolumn{1}{c|}{0.751}                                                                   & \multicolumn{1}{c|}{0.751}                                                                 & \multicolumn{1}{c|}{0.751}                                                                   & 0.751                                                                  \\ \hline
\multicolumn{1}{|c|}{\textbf{}}                                                                           & \multicolumn{4}{c|}{\textbf{C = 0.7, $\beta= -0.1$,  $\alpha= 0.1$}}                                                                                                                                                                                                                                                                                              \\ \hline
$\chi^2$                                                                                                  & \multicolumn{1}{c|}{32.889}                                                                  & \multicolumn{1}{c|}{32.897}                                                                & \multicolumn{1}{c|}{32.905}                                                                  & 32.914                                                                 \\ 
$\Omega_{de}$                                                                                       & \multicolumn{1}{c|}{0.731}                                                                   & \multicolumn{1}{c|}{0.731}                                                                 & \multicolumn{1}{c|}{0.731}                                                                   & 0.731                                                                  \\ \hline
\multicolumn{1}{|c|}{\textbf{}}                                                                           & \multicolumn{4}{c|}{\textbf{C = 0.7, $\beta= 0.1$,  $\alpha= -0.1$}}                                                                                                                                                                                                                                                                                              \\ \hline
$\chi^2$                                                                                                   & \multicolumn{1}{c|}{32.951}                                                                  & \multicolumn{1}{c|}{32.960}                                                                & \multicolumn{1}{c|}{32.968}                                                                  & 32.978                                                                 \\ 
$\Omega_{de}$                                                                                        & \multicolumn{1}{c|}{0.748}                                                                   & \multicolumn{1}{c|}{0.748}                                                                 & \multicolumn{1}{c|}{0.748}                                                                   & 0.748                                                                  \\ \hline
\multicolumn{1}{|c|}{\textbf{}}                                                                           & \multicolumn{4}{c|}{\textbf{C = 0.7, $\beta= -0.1$,  $\alpha= -0.1$}}                                                                                                                                                                                                                                                                                             \\ \hline
$\chi^2$                                                                                                   & \multicolumn{1}{c|}{32.951}                                                                  & \multicolumn{1}{c|}{32.729}                                                                & \multicolumn{1}{c|}{32.737}                                                                  & 32.746                                                                 \\ 
$\Omega_{de}$                                                                                       & \multicolumn{1}{c|}{0.748}                                                                   & \multicolumn{1}{c|}{0.728}                                                                 & \multicolumn{1}{c|}{0.728}                                                                   & 0.728                                                                  \\ \hline
\multicolumn{1}{|c|}{\textbf{}}                                                                            & \multicolumn{4}{c|}{\textbf{C = 1, $\beta= 0.1$,  $\alpha= 0.1$}}                                                                                                                                                                                                                                                                                                 \\ \hline
$\chi^2$                                                                                                   & \multicolumn{1}{c|}{32.561}                                                                  & \multicolumn{1}{c|}{32.568}                                                                & \multicolumn{1}{c|}{32.574}                                                                  & 32.581                                                                 \\ 
$\Omega_{de}$                                                                                        & \multicolumn{1}{c|}{0.764}                                                                   & \multicolumn{1}{c|}{0.762}                                                                 & \multicolumn{1}{c|}{0.762}                                                                   & 0.762                                                                  \\ \hline
\multicolumn{1}{|c|}{\textbf{}}                                                                            & \multicolumn{4}{c|}{\textbf{C = 1, $\beta= -0.1$,  $\alpha= 0.1$}}                                                                                                                                                                                                                                                                                                \\ \hline
$\chi^2$                                                                                                   & \multicolumn{1}{c|}{32.396}                                                                  & \multicolumn{1}{c|}{32.403}                                                                & \multicolumn{1}{c|}{32.409}                                                                  & 32.415                                                                 \\ 
$\Omega_{de}$                                                                                        & \multicolumn{1}{c|}{0.742}                                                                   & \multicolumn{1}{c|}{0.739}                                                                 & \multicolumn{1}{c|}{0.739}                                                                   & 0.739                                                                  \\ \hline
\multicolumn{1}{|c|}{\textbf{}}                                                                            & \multicolumn{4}{c|}{\textbf{C = 1, $\beta= 0.1$,  $\alpha= -0.1$}}                                                                                                                                                                                                                                                                                                \\ \hline
$\chi^2$                                                                                                   & \multicolumn{1}{c|}{32.473}                                                                  & \multicolumn{1}{c|}{32.481}                                                                & \multicolumn{1}{c|}{32.488}                                                                  & 32.495                                                                 \\ 
$\Omega_{de}$                                                                                        & \multicolumn{1}{c|}{0.762}                                                                   & \multicolumn{1}{c|}{0.762}                                                                 & \multicolumn{1}{c|}{0.759}                                                                   & 0.759                                                                  \\ \hline
\multicolumn{1}{|c|}{\textbf{}}                                                                            & \multicolumn{4}{c|}{\textbf{C = 1, $\beta= -0.1$,  $\alpha= -0.1$}}                                                                                                                                                                                                                                                                                               \\ \hline
$\chi^2$                                                                                                  & \multicolumn{1}{c|}{32.333}                                                                  & \multicolumn{1}{c|}{32.337}                                                                & \multicolumn{1}{c|}{32.344}                                                                  & 32.350                                                                 \\ 
$\Omega_{de}$                                                                                       & \multicolumn{1}{c|}{0.739}                                                                   & \multicolumn{1}{c|}{0.739}                                                                 & \multicolumn{1}{c|}{0.739}                                                                   & 0.739                                                                  \\ \hline
\end{tabular}
\caption{The results of matter perturbations data analysis for THDE model with interaction between matter and dark component in form $Q=H(\alpha\rho_m + \beta\rho_{de})$. It is interesting to note that for $C=1$ data are described better than for $C=0.7$ (for combined data analysis the situation is opposite).}
\label{Tab_9}
\end{table}

\begin{figure}
\centering
    \includegraphics[width=1\textwidth]{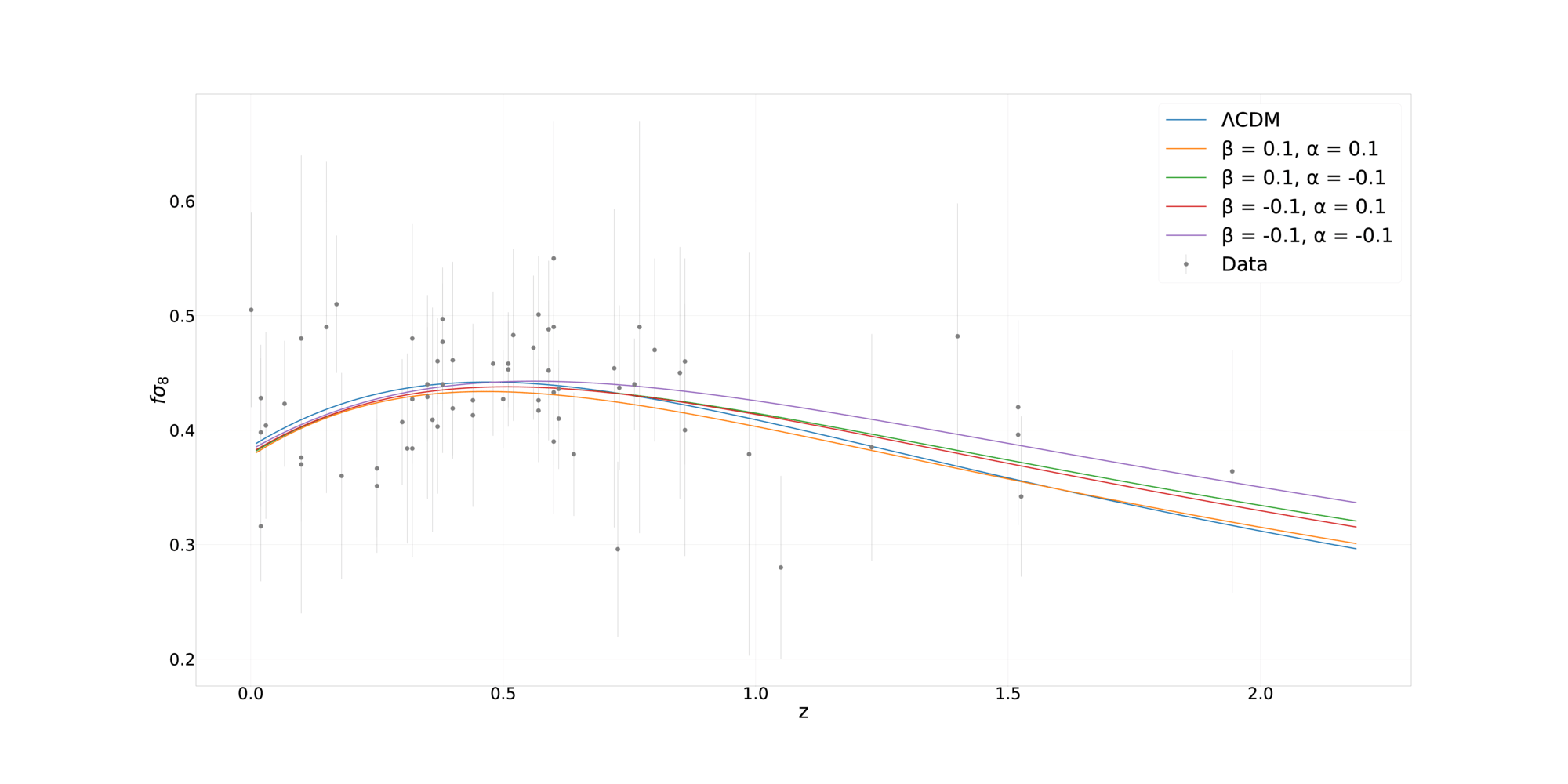}
    \caption{The $f\sigma_{8}$ dependence on redshift for THDE model with $C = 1, \gamma = 0.9$ for various $\alpha$ and $\beta$ in presence of interaction between matter and dark component.} 
    \label{fig_9}
\end{figure}

In conclusion of our analysis we consider interaction between matter and dark components. The following interaction is proposed 
\begin{equation}
\label{eq:9}
Q=H(\alpha\rho_m + \beta\rho_{de}),
\end{equation}
where $\alpha$ and $\beta$ are dimensionless constants. From hypothetical reasons it follows that the values of $\alpha$ and $\beta$ should be small. We investigate some variants (see Table \ref{Tab_7}), without of interaction results of the value $C=0.7$ turn out better. But the interaction ``saves'' the case $C=1$: for some $\alpha$ and $\beta$ concordance with observational data is quite acceptable. The likelihood contours can be seen on Fig. \ref{fig_8} for $C=0.7, \gamma=1.2$ and $\alpha=0.1,\beta=-0.1$.

Analysis of $f\sigma_8$ data shows that the likelihood of THDE models at various $\alpha$ and $\beta$ weakly depends on parameters depicted in table \ref{Tab_9}, although dependence $f\sigma_8(z)$ changes for various parameters depicted in figure \ref{fig_9}). 

\section{Conclusion}

We compared THDE models for some parameters $C$ and $\gamma$ to standard $\Lambda$CDM cosmology. Our consideration was based on observational data concerning distance-redshift diagram for SNe Ia, dependence of Hubble parameter from redshift, various relations from BAO data and evolution of matter density perturbations. The calculation of $\chi^2$ for various datasets in the case of fixed $C$ and $\gamma$ allows to estimate the likelihood of THDE models in comparison to each other and to standard cosmology. Analysis of Pantheon+ dataset demonstrates that THDE model satisfactorily describes data in a wide range of $C$ and $\gamma$. For $C>\geq 1$ the concordance with SNe Ia data are relatively better than for $\Lambda$CDM model. The best-fit value of $H_0$ lies within the narrow range of $72.3<H_0<72.7$ km/s/Mpc. For optimal $\Omega_{de}$ only at $C=0.7$ we obtained value close to the best-fit $\Omega_{de}$ in $\Lambda$CDM model, if $C=1.2$ $\Omega_{de} \rightarrow 0.73$. On the other hand, DESI-2024 data and Hubble parameter data  favor $C<1$. The best-fit value for $H_{0}$ from BAO and $H(z)$ data strongly depends on $C$ while $\Omega_{de}$ changes insignificantly. These results weakly depend on the parameter of non-additivity $\gamma$.  

The analysis of combined Pantheon+BAO+$H(z)$ dataset allowed us to define the optimal value of parameter $C$ ($C=0.7$ in our consideration) and rule out the case of $C\geq 1$ for which concordance with observational data gets significantly worse in comparison to $\Lambda$CDM model. Results for best-fit values $\Omega_{de}$ and $H_0$ insignificantly change with parameter $\gamma$. However inclusion of interaction between matter and holographic dark energy in simple form expands the allowable interval of $C$. For interaction in form $Q = H (\alpha\rho_m + \beta\rho_d)$, the case $C=1$ for some $\alpha$ and $\beta$ is valid from an observational viewpoint.

We note also that best-fit value for Hubble parameter $H_0$ from BAO and $H(z)$ data is larger for THDE in comparison with $\Lambda$CDM model. Therefore Hubble tension problem looks not so dramatic in holographic paradigm althouth puzzle remains unsolved.

\subsection*{Acknowlegment}

This research was supported by funds provided through the Russian Federal Academic Leadership Program “Priority 2030” at the Immanuel Kant Baltic Federal University (No. 075-02-2024-1430).

\end{document}